\documentclass[a4paper,11pt]{article}
\pdfoutput=1

\usepackage[utf8]{inputenc}
\usepackage[margin=1in]{geometry}
\usepackage[T1]{fontenc}
\usepackage{booktabs,multirow}
\usepackage{bbold}
\usepackage{subcaption}
\usepackage{amsmath}
\usepackage{mathtools,amssymb,braket,dsfont}
\usepackage{url}
\usepackage{graphicx}
\usepackage{multicol}
\usepackage{multirow}
\usepackage{float}
\usepackage[dvipsnames]{xcolor}
\usepackage{slashed}
\usepackage[final]{hyperref}
\def\linkcolor{cyan!70!black}
\hypersetup{
	colorlinks=true,
	linkcolor=\linkcolor,
	citecolor=\linkcolor,
	filecolor=\linkcolor,
	urlcolor=\linkcolor
}
\usepackage[numbers, sort&compress]{natbib}
\usepackage[utf8]{inputenc}
\usepackage[T1]{fontenc}
\usepackage{comment}
\usepackage{siunitx}
\usepackage{enumitem}
\usepackage{orcidlink}
\usepackage{cancel}

\newcommand{\beq}{\begin{equation}}
\newcommand{\eeq}{\end{equation}}
\newcommand{\bea}{\begin{eqnarray}}
\newcommand{\eea}{\end{eqnarray}}

\newcommand{\revised}[1]{#1}
\newcommand{\rerevised}[1]{#1}

\begin{document}

\hfill TTP25-042,  P3H-25-090

\mbox{}\vspace{1cm}

\begin{center}

	 	{\LARGE \bf \boldmath Hunting for Neutrino Texture Zeros with\\[2mm] Muon and Tau Flavor Violation}

        \vspace{1cm}

{\large Lorenzo Calibbi \orcidlink{0000-0002-9322-8076}}$\,^{\dag}$\,\footnote{\href{mailto:calibbi@nankai.edu.cn}{calibbi@nankai.edu.cn}.}, 
{\large Xiyuan Gao \orcidlink{0000-0002-1361-4736}}$\,^{\ddag}$\,\footnote{\href{mailto:xiyuan.gao@kit.edu}{xiyuan.gao@kit.edu}}, and 
{\large Man Yuan \orcidlink{0000-0002-8897-6328}}$\,^{\dag}$\,\footnote{\href{mailto:yuanman@mail.nankai.edu.cn}{yuanman@mail.nankai.edu.cn}} 

        \vspace{0.5cm}

{\emph{$^{\dag}$\,School of Physics, Nankai University, Tianjin 300071, China}}\\[2mm]
{\emph{$^{\ddag}$\,Institute for Theoretical Particle Physics, Karlsruhe Institute of Technology (KIT), Wolfgang-Gaede-Str.~1, D-76131 Karlsruhe, Germany}}

\end{center}
\vspace*{1cm}

\begin{abstract}
\noindent
We revisit the minimal type~II seesaw mechanism generating the Majorana neutrino mass matrix $M^{\nu}$, under the assumption that two entries of $M^{\nu}$ vanish. Such flavor structures are known as two-zero textures. Processes with charged lepton flavor violation (CLFV), absent in the Standard Model~(SM), can have sizable rates in this framework and are directly linked to the flavor structure of $M^{\nu}$. For each allowed two-zero texture, we quantify the predicted correlations among various CLFV observables using current neutrino oscillation data and show that they lead to distinctive patterns of CLFV processes that could be discriminated between at running and upcoming experiments. In addition, together with information from colliders, the sensitivity of these correlations to renormalization group (RG) effects could shed light on the potentially ultra-high scale where new dynamics (e.g.~some underlying flavor symmetry) give rise to the two-zero texture.
Furthermore, we find that certain zero textures, although not third-generation specific, can suppress $\mu\to e$ transitions while allowing the rate of the process $\tau\to \bar\mu ee$ to be within the future experimental sensitivity, even when the RG evolution is taken into account. 
The lowest possible cut-off scale of the effective theory, constructed by treating the two-zero flavor structure of $M^{\nu}$ as a CLFV spurion, can therefore reach $5-6$ TeV. 
Our results provide further motivation for searches for $\tau$ CLFV at Belle II, as probes of new physics complementary to MEG II and the upcoming Mu3e, COMET, and Mu2e experiments, as well as for collider searches for doubly charged scalar bosons.

\end{abstract}
\vspace*{0.5cm}


\thispagestyle{empty}

\setcounter{tocdepth}{1}

\newpage

\tableofcontents

\section{Introduction}
\setcounter{footnote}{0}

The Standard Model (SM) of particle physics has been tested with high precision across a wide range of experiments. Nevertheless, the observation of tiny neutrino masses indicates that the SM is not complete and new physics (NP) beyond it is required. NP particles and interactions generally mediate flavor changing transitions that are forbidden or highly suppressed within the SM, potentially leaving observable signals in high-precision and/or high-intensity frontier experiments. Among the latter, a prominent role is played by searches for processes with charged lepton flavor violation (CLFV) that are among the most powerful and cleanest low-energy probes of NP, as they are suppressed to a negligible level within the SM. These processes, which are presently the object of a robust experimental program aimed at increasing current sensitivities by several orders of magnitude~\cite{Calibbi:2017uvl}, are the main focus of the present work.

In addition to direct or indirect searches for new degrees of freedom beyond the SM, another widely explored direction to unveil new underlying dynamics is investigating possible internal correlations among the SM parameters. 
One fundamental feature of the SM is the rich flavor sector: we observe the existence of three families (or flavors) for each chiral fermion with given gauge quantum numbers, only differing in their masses. Although the flavor parameters, including fermion masses, mixing angles, and $CP$ violating (CPV) phases, are free parameters within the SM, they do not appear to be random. Rather, they mostly exhibit peculiar hierarchical patterns. The very existence of three families and the origin of the observed patterns of fermion masses and mixing is often referred to as the SM ``flavor puzzle''~\cite{Feruglio:2015jfa,Xing:2020ijf}, as the flavor sector has no fundamental principle explaining its structure within the SM.
In addition, a number of flavor parameters seem to obey special interrelationships. 
The existence of hidden flavor textures for the fermion mass matrices is therefore a well-motivated possibility. 

Already in 1977~\cite{Weinberg:1977hb, DeRujula:1977dmn, Wilczek:1977uh, Fritzsch:1977za}, it was noted that the phenomenological relation among the Cabibbo angle $\theta_c$ and quark masses $\theta_c=\arctan\sqrt{\frac{m_d}{m_s}}-\arctan\sqrt{\frac{m_u}{m_c}}$ may originate from quark mass matrices with vanishing $(1,1)$ entries:
\begin{equation}
    \label{FritzschTexture}
    M_D~=~\left( 
    \begin{matrix}
       0  & \sqrt{m_d m_s} \cr
       \sqrt{m_d m_s}  & m_s
    \end{matrix}
    \right), \quad 
    M_U~=~\left( 
    \begin{matrix}
       0  & \sqrt{m_u m_c} \cr
       \sqrt{m_u m_c}  & m_c
    \end{matrix}
    \right).
\end{equation}
Non-trivial flavor textures like the above one may also appear in the lepton sector. 
For instance, neutrino oscillation data hints at an approximate ``tribimaximal'' flavor mixing pattern~\cite{Fritzsch:1995dj, Harrison:2002er, Xing:2002sw, Harrison:2002kp, He:2003rm, Altarelli:2005yx}: $\theta_{12}=\arctan{\frac{1}{\sqrt{2}}}$, $\theta_{23}=\frac{\pi}{4}$, $\theta_{13} \ll \theta_{12},\theta_{23}$, where $\theta_{12}$, $\theta_{23}$, and $\theta_{13}$ are the mixing angles observed, respectively, in solar, atmospheric, and reactor neutrino oscillations. 
Since the absolute neutrino masses $m_{\nu_1}, m_{\nu_2},$ and $m_{\nu_3}$ are only constrained but not yet measured, the flavor texture of the neutrino mass matrix $M^\nu$ cannot be fully inferred. 
Despite this, texture zeros similar to those for quarks shown in Eq.~\eqref{FritzschTexture} are conjectured to arise for $M^{\nu}$~\cite{Frampton:2002yf, Xing:2002ta, Kageyama:2002zw, Xing:2002ap, Frigerio:2002fb, Desai:2002sz, Guo:2002ei}, in the basis where the charged lepton mass matrix is diagonal, as they introduce constraints that reduce the number of free parameters tightening the connection between flavor texture and measured quantities. 
In fact, current experimental measurements of neutrino oscillations allow up to two vanishing elements in $M^\nu$, and some of such one- and two-zero textures can be tested in the near future --- see, for instance, Refs.~\cite{Fritzsch:2011qv, Meloni:2014yea, Zhou:2015qua, Alcaide:2018vni,Singh:2019baq, Denton:2023hkx, Chauhan:2023faf, Treesukrat:2025dhd} for analyses published after the measurement of the reactor mixing angle.
For instance, even without knowledge of the absolute neutrino masses, future oscillation experiments such as JUNO~\cite{JUNO:2015zny} and DUNE~\cite{DUNE:2015lol} can shed light on certain two-zero textures, once the Dirac CPV phase is measured or constrained. 
\revised{Interestingly, a recent work~\cite{Borah:2025vtn} points out that for the case of a normal neutrino mass hierarchy, one of the two-zero textures allowed by NuFit-6.0 data~\cite{Esteban:2024eli} is in tension with the newly released JUNO 2025 data~\cite{JUNO:2025gmd}.}

Non-trivial flavor textures are generally employed for the interactions of NP fields as well. An extensively studied framework is minimal flavor violation (MFV)~\cite{Chivukula:1987py, DAmbrosio:2002vsn, Cirigliano:2005ck, Davidson:2006bd, Isidori:2012ts}, which assumes that, also in the NP sector, the $U(3)^5$ global flavor symmetry of the SM gauge sector is only broken by the SM Yukawas, $Y_u$, $Y_d$, and $Y_e$, acting as ``spurions''. A less restrictive flavor symmetry is $U(2)^5$, which only acts on the two lightest fermion families~\cite{Isidori:2012ts, Barbieri:2011ci, Barbieri:2012uh,Blankenburg:2012nx, Faroughy:2020ina, Allwicher:2023shc, Allwicher:2025bub}. Both frameworks are well-motivated since $U(3)^5$ and $U(2)^5$ are accidental symmetries of the SM gauge sector and, in particular, the latter one naturally addresses the observed fact that the first and second generations are, in first approximation, massless compared to third generation fermions. In addition, both paradigms ensure a sufficient protection of dangerous flavor-violating transitions between light fermions. 
Recently, Refs.~\cite{Allwicher:2023shc, Allwicher:2025bub} showed that for all $U(2)^5$-symmetric operators of the SM effective field theory (SMEFT), a TeV-scale cutoff is consistent with all present experimental bounds, given that the Wilson coefficients (WCs) of operators involving light fields and/or Higgs fields are mildly suppressed.

An interesting question then arises: are there any other flavor textures, for either renormalizable models or effective theories, that also suppress the light flavor transitions while allow third-generation or Higgs specific signals for near-future experiments? Since large neutrino mixing angles break both the $U(2)_{\ell}$ and $U(3)_{\ell}$ symmetry in the lepton sector, many other options for leptonic textures were also considered.
One approach is to explore the alternative (broken) subgroups of $U(3)_{\ell}\times U(3)_e$ that are distinct from $U(2)_{\ell}\times U(2)_{e}$~\cite{Heeck:2016xwg, Faroughy:2020ina, Greljo:2022cah, Greljo:2025mwj}. On the other hand, symmetries directly linked to the tribimaximal flavor mixing pattern are broadly studied as well, especially discrete symmetries that may originate from modular symmetries inspired by string compactification --- see Refs.~\cite{Feruglio:2008ht, Hagedorn:2009df, Feruglio:2009hu,  delAguila:2010vg,  Ding:2011gt, Altarelli:2012bn, Pascoli:2016wlt, Heinrich:2018nip, Kobayashi:2021pav, Kobayashi:2022jvy, Ding:2022nzn,  Bigaran:2022giz, Lichtenstein:2023iut, Chauhan:2023faf, Lichtenstein:2023vza, Palavric:2024gvu,  Moreno-Sanchez:2025bzz, Calibbi:2025fzi, Jangid:2025thp} for examples relating such symmetries to CLFV.
In some scenarios, $\mu\to e$ transition amplitudes vanish if certain discrete symmetries are exact --- see \cite{Calibbi:2025fzi} for a recent example --- with a flavor structure significantly differing from those restricted by the $U(2)^5$ ansatz. 
This fact indicates that the possible flavor structures for the TeV-scale NP are highly diverse. Furthermore, beyond the extensively explored landscape of flavor symmetries, a wide range of equally predictive textures may still remain uncovered. 

The purpose of this work is to study a set of more general flavor textures that suppress $\mu\to e$ transitions and predict correlated CLFV processes across all three generations. In particular, we focus on the above-mentioned texture zeros in the neutrino mass matrix that have been first introduced in~\cite{Frampton:2002yf}.
Working within the minimal type~II seesaw model~\cite{Magg:1980ut, Lazarides:1980nt, Schechter:1980gr, Mohapatra:1980yp, Gelmini:1980re} as a benchmark, we find that certain textures that do not entail enhanced flavor symmetries can also be consistent with $5-10$~TeV new physics. 
We choose type~II seesaw model because it is \revised{one of} the simplest UV-complete neutrino mass model (only requiring the introduction of a new scalar field) and it is therefore highly predictive: (i)~CLFV interactions are directly linked to the neutrino mass matrix without involving further free parameters, (ii)~they can be sizable while still leading to tiny neutrino masses through a small lepton number breaking parameter (with no need to extend the model). 

Specifically, the model extends the SM by adding an $SU(2)_L$ triplet field $\Delta$ with non-zero hypercharge. In the limit of a small breaking of the lepton number, the mass of $\Delta$ can be as low as the TeV scale. 
As mentioned, the $\Delta$-lepton Yukawa coupling $Y_{\Delta}$ is proportional to the neutrino mass matrix $M^{\nu}$, which breaks the chiral $U(2)_{\ell}$ flavor symmetry because the approximate tribimaximal flavor mixing pattern is not third-generation specific. 
Restoring the $U(2)_{\ell}$ symmetry and suppressing $\mu\to e$ transitions is only possible by taking the limit $Y_{\Delta}\to 0$ that, however, pushes the effective cutoff scale to high values and suppresses all other CLFV operators as well.
In this work, we find that the limits of a heavy $\Delta$ or preserved $U(2)_{\ell}$ are not necessary conditions to evade the tight $\mu\to e$ constraints. An alternative way out is requiring $M^{\nu}$ to exhibit the above-mentioned texture zeros. This conclusion holds even when one-loop matching and running of the CLFV WCs are included. 

If the zero textures of $M^{\nu}$ are protected by (generalized) flavor symmetries, the minimal seesaw model needs extensions in general --- see, for instance, Refs.~\cite{Grimus:2004hf, Grimus:2004az, Hirsch:2007kh, Dev:2011jc, Fritzsch:2011qv, Araki:2012ip, Dev:2014dla, GonzalezFelipe:2014zjk, Lamprea:2016egz, Linster:2018avp, Borgohain:2018lro, Verma:2018lro, Zhang:2019ngf, Bjorkeroth:2019rat, Linster:2020fww, Barreiros:2022aqu, Ding:2022nzn, Ding:2022aoe,  Ding:2023htn, Rocha:2024twm, Fang:2024qtx, Rocha:2025ade, Jiang:2025psz}. However, we find that within minimal type~II seesaw, even if the renormalization group (RG) evolution from a high scale $\Lambda_\text{UV}$ is included, $\mu\to e$ can still be sufficiently suppressed. 
As a result, the landscape of type~II seesaw becomes more ``flavorful'': besides $\mu\to \bar e ee$, $\mu\to e\gamma$, and $\mu-e$ conversion in nuclei, $\tau\to\overline{\mu}ee$, the process of $\tau$ decaying to one muon and two same-sign electrons, could also provide the first signal for CLFV within type~II seesaw. On the other hand, if this $\tau$ CLFV decay is observed, the $\mu\to e$ processes must also be sizable and are guaranteed to be detected at running or upcoming experiments. The correlations of multiple CLFV processes and their specific patterns can give a hint on the underlying flavor structure of $Y_{\Delta}$, and perhaps also shed light on the scale $\Lambda_\text{UV}$ where this structure originates. 
Therefore, we provide a further theoretical motivation to search for $\tau$ CLFV in high-intensity frontier experiments such as Belle~II, as a probe complementary to the currently running MEG~II and the upcoming Mu3e, COMET, and Mu2e experiments. 

Some works in the literature have partly explored similar ideas, in particular~\cite{Chun:2003ej, Akeroyd:2009nu, Ardu:2024bua,Kitabayashi:2020ajn}. 
Refs.~\cite{Chun:2003ej, Akeroyd:2009nu} are early studies on CLFV within type~II seesaw that also discussed the conditions on $Y_{\Delta}$ required to suppress $\mu\to e$ transitions, which additionally link CLFV processes to neutrino oscillation parameters. However, the crucial one-loop enhancement for $\mu\to \bar e ee$ was not included. In the context of a general bottom-up effective field theory~(EFT) analysis, Ref.~\cite{Ardu:2024bua}
studied the full RG running of the CLFV WCs (hence capturing all one-loop effects) and mapped them to UV-complete models including type~II seesaw. In this context, 
they considered the one-zero texture with $M^\nu_{e\mu} = 0$ to suppress $\mu\to e$ transitions and
presented correlations between $\mu\to \bar e ee$ and the other CLFV decay rates for a benchmark choice of the neutrino mixing parameters. 
Nevertheless, we find that the correlations with $\tau$ CLFV feature large variance when the uncertainty of the neutrino parameters is taken into account. Predictive patterns of CLFV observables can only be identified within more constraining textures for $Y_{\Delta}$, e.g.~with two vanishing entries. 
To the best of our knowledge, two-zero textures for $Y_{\Delta}$ were only considered in Ref.~\cite{Kitabayashi:2020ajn}, where the one-loop enhanced contribution to $\mu\to \bar e ee$ was again omitted and other loop-induced processes, such as $\mu\to e\gamma$, were not included in the analysis. In addition, all previous works assumed that the zero textures for $Y_{\Delta}$ are defined exactly at the scale $m_{\Delta}$. If this assumption is relaxed, corrections originated from the RG evolution from $\Lambda_{\text{UV}}$ to $m_{\Delta}$ have yet to be  quantified.

The paper is structured as follows. In Section~\ref{modelsection}, we first revisit the Lagrangian of minimal type~II seesaw and summarize the relevant effective operators up to one-loop matching. In Section~\ref{texturesection}, we review the seven realistic two-zero textures for the Majorana neutrino mass matrix $M^{\nu}$, and apply them to the Yukawa matrix $Y_{\Delta}$. We then analyze how the current CLFV experiments constrain the effective cutoff scale for each texture and briefly compare our results with other patterns predicted by $U(2)$, $A_4$, and $Z_3$ flavor symmetries. In the following Section~\ref{RGstability}, the stability of the two-zero textures under the renormalization group (RG) running is examined in both regions above and below the scale of $\Delta$. In Section~\ref{phenosection}, we first analyze the correlation among various $\mu$ and $\tau$ CLFV processes and then discuss the absolute transition rates. After that, we quantify the RG evolution corrections to the predicted ratios of the branching ratios (BRs), and explain why they can shed light on scales far higher than that of $\Delta$. We also comment on complementary information that could be obtained at high-energy colliders. Finally, our findings are summarized in Section~\ref{conclu}. In the appendix we give more technical details on our calculations. 
\revised{In Appendix~\ref{loopfunction}, we provide the loop functions related to the penguin diagram. 
The explicit expressions of the dependence of $M^{\nu}$ with two-zero textures on the neutrino oscillation parameters, and present benchmark matrices for each pattern, are collected in Appendix~\ref{expressions}. Appendix~\ref{overall} presents more details on the RG running.}

\section{Type~II seesaw and the low energy theory}
\label{modelsection}

The complete Lagrangian of the type~II seesaw model reads (see e.g.~\cite{Abada:2007ux, FileviezPerez:2008jbu} for reviews): 
\begin{equation}
\label{minimalTypeII}
\begin{aligned}
    \mathcal{L}~=&~\mathcal{L}_{\text{SM}}
    +\text{Tr}\left[(D_{\mu}\Delta)^{\dagger}(D^{\mu}\Delta)\right]-m_{\Delta}^2\text{Tr}(\Delta\Delta^{\dagger})
    -V^{(4)}(H,\Delta)\\
    &-(Y_{\Delta}\overline{\ell_L}\Delta i\tau_2 \ell_L^c -\mu_\Delta H^T i\tau_2 \Delta H+\text{h.c.})\,,\\
\end{aligned}
\end{equation}
where $\ell_L$ is the left-handed lepton doublet (with $\ell_L^c = i\gamma_2\gamma_0\ell_L$ denoting the charge-conjugated field), $H$ is the SM Higgs doublet, and $\Delta$ is the new ${SU}(2)_{L}$ scalar triplet:
\begin{equation}
\label{Delta}
    \Delta \equiv 
    \left(\begin{array}{cc}
       \Delta^-/\sqrt{2}  & \Delta^0 \\
        \Delta^{--} & -\Delta^-/\sqrt{2}
    \end{array}\right)\,.
\end{equation}

In the above Lagrangian, $Y_{\Delta}$ is a $3\times3$ complex symmetric matrix, whose flavor indices are not shown explicitly, and $\mu$ is the trilinear coupling between the triplet and the SM Higgs doublet. 
%
The quartic self-interactions involving $\Delta$ and $H$ in $V^{(4)}(H,\Delta)$ have the general form:
\begin{equation}
     V^{(4)}(H,\Delta) ~=~ \lambda_1 \left[ \text{Tr}(\Delta^{\dagger}\Delta)\right]^2+\lambda_2 \text{Det}(\Delta^{\dagger}\Delta) + \lambda_3 (H^{\dagger}H) \text{Tr}(\Delta^{\dagger}\Delta)+\lambda_4 (H^{\dagger}\sigma_iH)\text{Tr}(\Delta^{\dagger}\sigma_i\Delta)\,.
\end{equation}

Once the scalar doublet $H$ develops a vacuum expectation value (VEV) $v \simeq 246~\text{GeV}$ by spontaneous symmetry breaking, this also induces a VEV $v_{\Delta}$ for the triplet $\Delta$, from which the Majorana neutrino masses matrix arises: 
\begin{equation}
\label{massmatrix}
    M^{\nu}~=~ \sqrt{2} Y_{\Delta}v_{\Delta}~=~\mu_\Delta  Y_{\Delta} \frac{v^2}{m_{\Delta}^2}\,. 
\end{equation}
As customary, $M^{\nu}$ is diagonalized by a unitary transformation~\cite{Pontecorvo:1957cp, Maki:1962mu}:
\begin{equation}
M^{\nu} ~=~ U \left( \begin{matrix} m_{\nu_1} & 0 & 0 \cr 0 & m_{\nu_2} & 0 \cr 0 & 0 & m_{\nu_3} \end{matrix} \right) U^{\rm T}\,,
\label{eq:mnu}
\end{equation}
where the mixing matrix $U$ can be parametrized by three mixing angles and three CPV phases:
\begin{eqnarray}
U = \left( \begin{matrix} c^{}_{13} c^{}_{12} & c^{}_{13} s^{}_{12} & s^{}_{13} e^{-{\rm i}\delta} \cr -s_{12}^{} c_{23}^{} - c_{12}^{} s_{13}^{} s_{23}^{} e^{{\rm i}\delta}_{} & + c_{12}^{} c_{23}^{} - s_{12}^{} s_{13}^{} s_{23}^{} e^{{\rm i}\delta}_{} & c_{13}^{} s_{23}^{} \cr + s_{12}^{} s_{23}^{} - c_{12}^{} s_{13}^{} c_{23}^{} e^{{\rm i}\delta}_{} & - c_{12}^{} s_{23}^{} - s_{12}^{} s_{13}^{} c_{23}^{} e^{{\rm i}\delta}_{} & c_{13}^{} c_{23}^{} \end{matrix} \right).
\left( \begin{matrix} e^{i\rho} & 0 & 0 \cr 0 & e^{i\sigma} & 0 \cr 0 & 0 & 1 \end{matrix} \right),
\label{eq:pmns}
\end{eqnarray}
with $c^{}_{ij} \equiv \cos \theta^{}_{ij}$ and $s^{}_{ij} \equiv \sin \theta^{}_{ij}$. Throughout the paper, we work in the basis with a flavor-diagonal charged-lepton mass matrix, hence $U$ represents the matrix that relates neutrino flavor and mass eigenstates.

In the limit $\mu_\Delta=0$, the lepton number $L$ is conserved, that is, the Lagrangian in Eq.~\eqref{minimalTypeII} is invariant under the global $U(1)$ symmetry:
\begin{equation}
    \Delta\to\Delta e^{-2i\theta_L}, \quad \ell_L\to \ell_L e^{i\theta_L},  \quad H\to H\,. 
\end{equation}
In other words, if $\mu_\Delta\neq 0$, thus for non-vanishing Majorana neutrino mass terms in $M^{\nu}$, $L$ is explicitly broken. Neutrino masses can thus be suppressed by a small value of $\mu_\Delta$ --- a situation that can be regarded as ``natural'' since it leads to the enhancement of the symmetry of the theory~\cite{tHooft:1979rat}. As a consequence, large Yukawa couplings, $Y_{\Delta}\sim\mathcal{O}(1)$, and/or a relatively low-energy triplet mass, $m_{\Delta}\sim \mathcal{O}(\text{TeV})$, are not in conflict with the observed smallness of neutrino masses. This opens up the possibility of testing the theory searching for $L$-conserving CLFV processes, see e.g.~\cite{Abada:2007ux,Dinh:2012bp}. 

\begin{figure}[!t]
    \centering
    \includegraphics[width=1\linewidth]{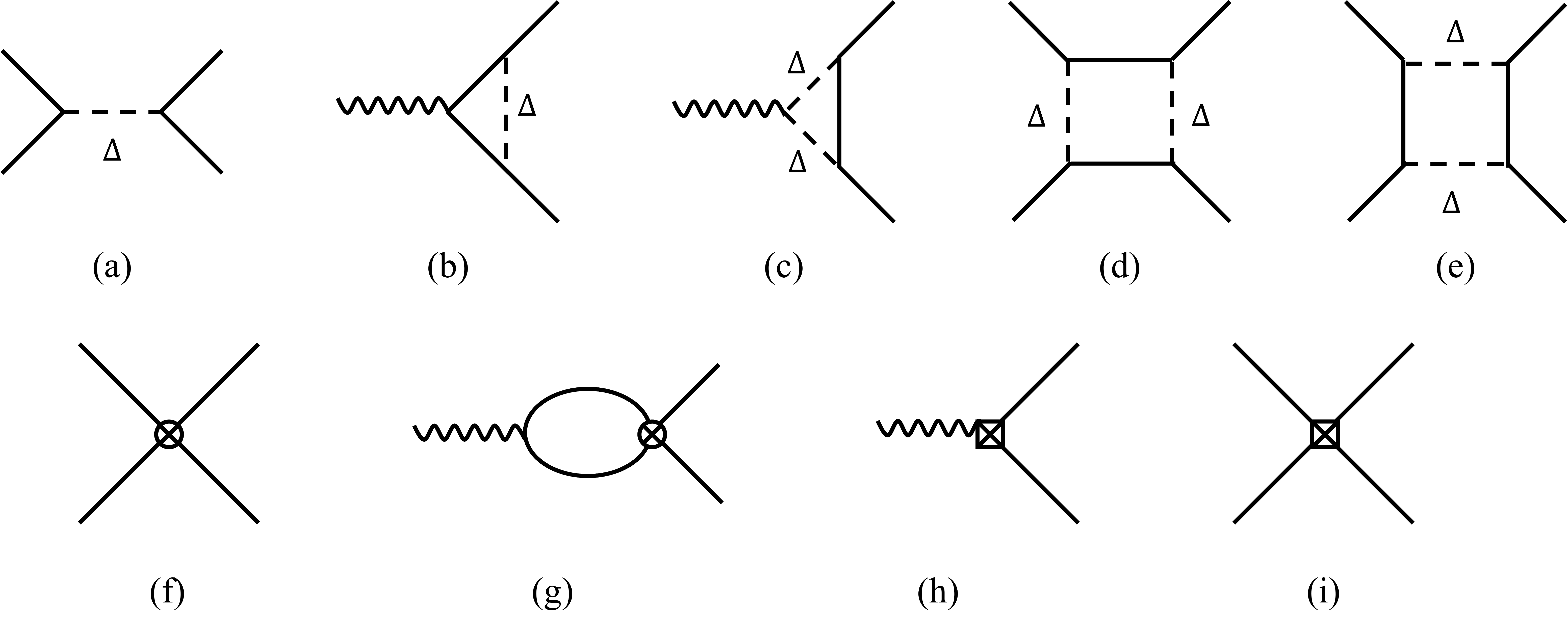}
    \caption{(a)-(e): Feynman diagrams contributing to CLFV processes within the full type~II seesaw model. (f)-(i): Corresponding Feynman diagrams in the effective theory at low-energy scales, $q^2\ll m_{W}^2$. We do not apply the on-shell condition so a pair of
    quarks or charged leptons can be attached to the external gauge boson lines. Self-energy diagrams are not included for simplicity.}
    \label{Feynmandiagrams}
\end{figure}

The doubly-charged state of the triplet directly contributes to processes such as $\mu\to \bar e ee$ via the diagram (a) of Figure~\ref{Feynmandiagrams}. 
Upon integrating out $\Delta$, this amplitude matches to the following low-energy CLFV operators:
\begin{equation}
\label{eftTree}
\delta\mathcal{L}^{\text{d=6}}_{\text{tree}}~=~ \frac{1}{2m_{\Delta}^2} Y_{\Delta jl}Y_{\Delta ik}^{*}(\overline{\ell_{i}}\gamma_{\mu}P_L\ell_{j}) (\overline{\ell_{k}}\gamma^{\mu} P_L\ell_{l})\,,
\end{equation}
where $\ell_i=e_{L\,i}+e_{R\,i}$ with $e_{L}$ ($e_{R}$) denoting the left-handed (right-handed) charged leptons and $i,j,k,l=e,\mu, \tau$ being flavor indices. 
According to \cite{Li:2022ipc}, the WCs of all the other dimension-six operators originating from tree-level matching are proportional to $\mu_\Delta$ and are therefore suppressed when $\mu_\Delta\ll m_{\Delta}$. 
At one-loop level, quark flavor-violating operators can arise but all their WCs also depend on $\mu_\Delta$~\cite{Li:2022ipc}. As a consequence, constraints from lepton flavor universality (LFU) tests in flavor-changing hadronic decays, such as the bounds from $R(K^{(*)})\equiv\Gamma(B\to K^{(*)}\mu^+\mu^-)/\Gamma(B\to K^{(*)}e^+e^-)$~\cite{LHCb:2022qnv, LHCb:2022vje} , are automatically evaded.

On the other hand, a set of new CLFV operators that arise at one-loop level can be sizable. They are complementary to $\delta\mathcal{L}^{\text{d=6}}_{\text{tree}}$ because they feature different flavor structures. 
At low energy scales $q^2\ll m_W^2$, we have:
\begin{equation}
\label{eft}
\begin{aligned}
    \delta\mathcal{L}^{\text{d=6}}_{\text{FV}}~=&~\mathcal{C}_{\text{dipole}}^{ik} \left(\overline{\ell_{i}}\sigma^{\mu\nu} P_R \ell_{k}\right) F_{\mu\nu}+\mathcal{C}_{\text{penguin}}^{ik}(\overline{\ell_{i}}\gamma_{\mu}P_L\ell_{k}) \left(\overline{\ell_{l}}\gamma^{\mu} \ell_{l}+\frac{2}{3}\overline{u_{l}}\gamma^{\mu}  u_{l}-\frac{1}{3}\overline{d_{{l}}}\gamma^{\mu} d_{l}\right)\\
    ~&+~ \mathcal{C}_{\text{box}}^{ijkl}(\overline{\ell_{i}}\gamma_{\mu}P_L\ell_{j}) (\overline{\ell_{k}}\gamma^{\mu} P_L\ell_{l})\,, 
\end{aligned}
\end{equation}
where 
\begin{equation}
\label{eftWilson}
    \begin{aligned}
        \mathcal{C}_{\text{penguin}}^{ik}~=&~\frac{e^2}{12\pi^2 m_{\Delta}^2} Y_{\Delta\,ij}Y_{\Delta\,kj}^{*}\left(\frac{1}{12}+f\left(\frac{q^2}{m_{\Delta}^2}\,,\frac{m_j^2}{m_{\Delta}^2}\right)\right),\\
        \mathcal{C}_{\text{dipole}}^{ik}~=&~\frac{3e }{64 \pi^2 m_{\Delta}^2} (Y_{\Delta}Y_{\Delta}^{\dagger})_{ik} m_k\,,\\
        \mathcal{C}_{\text{box}}^{ijkl}~=&~ -\frac{1}{32\pi^2m_{\Delta}^2}(Y_{\Delta}Y_{\Delta}^{\dagger})_{ij}(Y_{\Delta}Y_{\Delta}^{\dagger})_{kl}-\frac{1}{8\pi^2m_{\Delta}^2}(Y_{\Delta}Y_{\Delta}^{\dagger})_{il}(Y_{\Delta}Y_{\Delta}^{\dagger})_{kj}\,.\\
    \end{aligned}
\end{equation}
The above one-loop matching is also discussed in~\cite{Dinh:2012bp, Li:2022ipc, Ardu:2023yyw}. The corresponding Feynman diagrams of the full renormalizable theory are illustrated in the panels (b)-(e) of Figure~\ref{Feynmandiagrams}. Notice that we do not use the standard operator basis (see e.g.~\cite{Jenkins:2017jig}) of the low-energy effective field theory (LEFT), in order to keep a clear correspondence between operators and diagrams of Figure~\ref{Feynmandiagrams}.

As we do not enforce the on-shell condition for the external gauge bosons, a fermion line (hence a pair of quarks or charged leptons) can be attached.
Both diagram (b) and (c) contribute to $\mathcal{C}_{\text{penguin}}$ and the explicit form for the corresponding loop function $f$ is shown in Appendix~\ref{loopfunction}. In the limit $q^2 \lesssim m_j^2 \ll m_{\Delta}^2$ (corresponding to a $\mu$ decay through $\mu$ or $\tau$ in the loop), this reads~\cite{Raidal:1997hq}:
\begin{equation}
\label{loopf}
    f\left(\frac{-q^2}{m_{\Delta}^2},\frac{m_j^2}{m_{\Delta}^2}\right)~=~2\log\left(\frac{m_j}{m_{\Delta}}\right)+\frac{5}{3}\,, \qquad m_j=m_{\mu}~\text{or}~m_{\tau}\,.
\end{equation}
The large mass hierarchy ensures that varying $m_{\Delta}$ from 1 to 10 TeV does not significantly change the numerical value of Eq.~\eqref{loopf}, so for simplicity we always take $m_{\Delta}=3$ TeV as a benchmark. In the view of the effective theory, the $\log$ function contained in $\mathcal{C}_{\text{penguin}}$ encodes the RG running of the tree-level WCs in diagram (f). This running evolves from $m_{\Delta}$ to $m_{\mu}$ or $m_{\tau}$, and its anomalous dimension can be calculated from diagram (g) of Figure~\ref{Feynmandiagrams}. Diagrams (h) and (i) cancel the divergent part of diagram (g) as counter terms. Their finite parts are matched from diagram (b)-(e) and lead to $\mathcal{C}_{\text{dipole}}$ and $\mathcal{C}_{\text{box}}$ in Eq.~\eqref{eftWilson}.

We calculate the tree-level and one-loop contributions to CLFV processes inserting the above coefficients into the expressions provided in Ref~\cite{Kuno:1999jp} and show the results in Table~\ref{3body}. The one-loop BRs are calculated in the limit that the tree-level amplitudes are zero (as the former would be subdominant otherwise), so interference terms are not displayed. For simplicity, we take $m_{j}=m_{\tau}$ to evaluate Eq.~\eqref{loopf}, which is a good approximation for all one-loop $\tau$ decay BRs.\footnote{Instead, for what concerns $\text{BR}(\mu\to \bar e ee)$, the expression of the table should be modified by adding the term $\left(25\left|(Y_{\Delta}Y_{\Delta}^{\dagger})_{ee} \right|^2+26\,\text{Re}\left[(Y_{\Delta}Y_{\Delta}^{\dagger})_{ee} \right]+10 \right) \left|(Y_{\Delta}Y_{\Delta}^{\dagger})_{e\mu}  \right|^2$ to account for $\mu$ running in the loop, which enhances the log term by a factor $\approx 1.8$.} 
Moreover, we define:
\begin{equation}
\label{calY}
    \begin{aligned}
    (\mathcal{Y}_{\Delta}\mathcal{Y}_{\Delta}^{\dagger})_{\mu\mu}~=~&(Y_{\Delta}Y_{\Delta}^{\dagger})_{\mu\mu}-4\frac{(Y_{\Delta}Y_{\Delta}^{\dagger})_{e\mu}(Y_{\Delta}Y_{\Delta}^{\dagger})_{\tau\mu}^*}{(Y_{\Delta}Y_{\Delta}^{\dagger})_{e\tau}}\,, \\
    (\mathcal{Y}_{\Delta}\mathcal{Y}_{\Delta}^{\dagger})_{ee}~=~& (Y_{\Delta}Y_{\Delta}^{\dagger})_{ee}-4\frac{(Y_{\Delta}Y_{\Delta}^{\dagger})_{\mu e}(Y_{\Delta}Y_{\Delta}^{\dagger})_{\tau e}^*}{(Y_{\Delta}Y_{\Delta}^{\dagger})_{\mu \tau}}\,,\\
    \end{aligned}
\end{equation}
in order to simplify the expressions of box diagram contributions.

\begin{table}[t!]
\renewcommand\arraystretch{2.5}
\resizebox{\textwidth}{!}{
    \begin{tabular}{l|c | c | c }
    \hline
       Observable  & Tree-level ($\times m_{\Delta}^4G_F^2 $) & One-loop ($\times 10^5 \,m_{\Delta}^4G_F^2$)  & 90~\% CL UL \\
    \hline
      $\text{BR}(\mu\xrightarrow{}\overline{e}ee)$   & $0.25\left|Y_{\Delta \mu e}Y_{\Delta ee}\right|^2$ & $\left(25\left|(Y_{\Delta}Y_{\Delta}^{\dagger})_{ee}  \right|^2+19\, \text{Re}\left[(Y_{\Delta}Y_{\Delta}^{\dagger})_{ee} \right]+5.8\right)  \left|(Y_{\Delta}Y_{\Delta}^{\dagger})_{e\mu}  \right|^2 $ & $1.0\times 10^{-12}$~\cite{SINDRUM:1987nra}\\
      $\text{BR}(\tau\xrightarrow{}\overline{e}ee)$   & $0.085\left|Y_{\Delta \tau e}Y_{\Delta ee}\right|^2$ & $\left(25\left|(Y_{\Delta}Y_{\Delta}^{\dagger})_{ee}  \right|^2+19\, \text{Re}\left[(Y_{\Delta}Y_{\Delta}^{\dagger})_{ee} \right]+6.3 \right)  \left|(Y_{\Delta}Y_{\Delta}^{\dagger})_{e\tau}  \right|^2 $ & $2.5\times 10^{-8}$~\cite{Belle-II:2025urb}\\
      $\text{BR}(\tau\xrightarrow{}\overline{e}e\mu)$   & $0.043\left|Y_{\Delta \tau e}Y_{\Delta e \mu}\right|^2$ & $\left(0.085\left|(\mathcal{Y}_{\Delta}\mathcal{Y}_{\Delta}^{\dagger})_{ee}  \right|^2+0.32\,\text{Re}\left[{(\mathcal{Y}_{\Delta}\mathcal{Y}_{\Delta}^{\dagger})}_{ee} \right]+0.76\right)  \left|(Y_{\Delta}Y_{\Delta}^{\dagger})_{\mu\tau}  \right|^2 $ & $1.6\times 10^{-8}$~\cite{Belle-II:2025urb} \\
      $\text{BR}(\tau\xrightarrow{}\overline{\mu}ee)$   & $0.043\left|Y_{\Delta \tau\mu}Y_{\Delta ee}\right|^2$ &  $25\left|(Y_{\Delta}Y_{\Delta}^{\dagger})_{e\mu}\right|\left|(Y_{\Delta}Y_{\Delta}^{\dagger})_{e\tau}\right|^2$ & $1.5\times 10^{-8}$~\cite{Hayasaka:2010np}\\
      $\text{BR}(\tau\xrightarrow{}\overline{e}\mu\mu)$   & $0.043\left|Y_{\Delta \tau e}Y_{\Delta \mu\mu}\right|^2$ & 25$\left|(Y_{\Delta}Y_{\Delta}^{\dagger})_{\mu e}\right|\left|(Y_{\Delta}Y_{\Delta}^{\dagger})_{\mu\tau}\right|^2$ & $1.3\times 10^{-8}$~\cite{Belle-II:2025urb}\\
      $\text{BR}(\tau\xrightarrow{}\overline{\mu}\mu e)$   & $0.085\left|Y_{\Delta \tau\mu}Y_{\Delta \mu e}\right|^2$ & $\left(0.085\left|(\mathcal{Y}_{\Delta}\mathcal{Y}_{\Delta}^{\dagger})_{\mu\mu}  \right|^2+0.32\,\text{Re}\left[(\mathcal{Y}_{\Delta}\mathcal{Y}_{\Delta}^{\dagger})_{\mu\mu} \right]+0.63\right) \left|(Y_{\Delta}Y_{\Delta}^{\dagger})_{e\tau}  \right|^2 $ & $2.4\times 10^{-8}$~\cite{Belle-II:2025urb}\\
      $\text{BR}(\tau\xrightarrow{}\overline{\mu}\mu\mu)$   & $0.043\left|Y_{\Delta \tau\mu}Y_{\Delta \mu\mu}\right|^2$ & $\left(4.3\left|(Y_{\Delta}Y_{\Delta}^{\dagger})_{\mu\mu}  \right|^2+3.2 \,\text{Re}\left[(Y_{\Delta}Y_{\Delta}^{\dagger})_{\mu\mu} \right]+0.93\right)  \left|(Y_{\Delta}Y_{\Delta}^{\dagger})_{\mu\tau}  \right|^2 $ & $1.9\times 10^{-8}$~\cite{Belle-II:2024sce}\\
      \hline
    $\text{BR}(\mu\xrightarrow[]{}e\gamma)$ & 0 &$98 \left|(Y_{\Delta}Y_{\Delta}^{\dagger})_{e\mu}\right|^2$  & $1.5\times 10^{-13}$~\cite{MEGII:2025gzr}\\
    $\text{BR}(\tau\xrightarrow[]{}e\gamma)$ & 0 & $17\left| (Y_{\Delta}Y_{\Delta}^{\dagger})_{e\tau}\right|^2 $ & $3.3\times 10^{-8}$~\cite{BaBar:2009hkt}\\
    $\text{BR}(\tau\xrightarrow[]{}\mu\gamma)$ & 0 & $17 \left|(Y_{\Delta}Y_{\Delta}^{\dagger})_{\mu\tau}\right|^2$ & $4.2\times 10^{-8}$~\cite{Belle:2021ysv}\\
    \hline
    $\text{CR}(\mu\text{Au}\xrightarrow[]{} e \text{Au})$ & 0 &  $71 \left|(Y_{\Delta}Y_{\Delta}^{\dagger})_{e\mu}\right|^2$ & $7.0\times 10^{-13}$~\cite{SINDRUMII:2006dvw}\\
    $\text{CR}(\mu\text{Al}\xrightarrow[]{} e \text{Al})$ & 0 &  $32 \left|(Y_{\Delta}Y_{\Delta}^{\dagger})_{e\mu}\right|^2$ & ---\\
    \hline
    \end{tabular}}
    \caption{Branching ratios of various CLFV processes in minimal type~II seesaw. 
    \revised{For illustration, the one-loop BRs are shown in the limit that the tree-level amplitudes vanish, so interference terms are not displayed. However, in the following numerical analysis, these contributions are included.}
    The last column displays the current experimental 90~\% confidence level (CL) upper limits (UL). The notation $(\mathcal{Y}_{\Delta}\mathcal{Y}_{\Delta}^{\dagger})_{ee/\mu\mu}$ is defined in Eq.~\eqref{calY}.
    \revised{We take $m_{\Delta}=3$ TeV to evaluate the logarithmic function in Eq.~\eqref{loopf}. However, varying $m_{\Delta}$ from 1 to 10 TeV does not significantly change the numerical values displayed.}}
    \label{3body}
\end{table}

It is \revised{worth remarking} that the Higgs and the $W$ and $Z$ gauge bosons, as well as the top quark, do not appear in the loop diagrams of Figure~\ref{Feynmandiagrams}. Hence, one can directly match the UV theory to the LEFT, instead of the SMEFT as an intermediate step, which is what we did above. However, being an $SU(2)_L$ triplet, $\Delta$ also couple to the $W$, $Z$, and $H$ bosons, thus giving rise to the following SMEFT operators that inducing flavor-violating $Z$ and $H$ decays: 
\begin{equation}
\label{EFTZpole}
\begin{aligned}
    \delta\mathcal{L}^{\text{d=6}}_{\text{Z,H-FV}}~=&~
    \mathcal{C}_{eW}(\overline{\ell_L}\sigma^{\mu\nu}e_R)\tau^IHW_{\mu\nu}^I+
    \mathcal{C}_{eB}(\overline{\ell_L}\sigma^{\mu\nu}e_R)HB_{\mu\nu}+
    \mathcal{C}_{eH}(H^{\dagger}H)(\overline{\ell_L}e_RH)\\
    &~+\mathcal{C}^{(3)}_{H\ell}(H^{\dagger}\tau^Ii\overset{\leftrightarrow}{D}_{\mu} H)(\overline{\ell_L}\tau^I\gamma^{\mu}\ell_L)+
    \mathcal{C}^{(1)}_{H\ell}(H^{\dagger}i\overset{\leftrightarrow}{D}_{\mu} H)(\overline{\ell_L}\gamma^{\mu}\ell_L)\,.
\end{aligned}
\end{equation}
At the scale $q^2 = m^2_W$, these operators are then matched to LEFT operators and their contributions are already included in the coefficients of dipole and four-fermion operators shown in Eq.~(\ref{eftTree}) and Eq.~(\ref{eft}). Compared to the CLFV decays of $\mu$ and $\tau$ leptons, the $Z$-pole observables such as $Z \to \bar \ell_i \ell_j$ currently provide weaker constraints on the set of coefficients \{$\mathcal{C}_{eW},\mathcal{C}_{eB},\mathcal{C}_{eH},\mathcal{C}^{(3)}_{H\ell}, \mathcal{C}^{(1)}_{H\ell}$\}~\cite{Calibbi:2021pyh}, so we do not include them in the following analysis. 
We also do not include the operator $(H^{\dagger}i\overset{\leftrightarrow}{D}_{\mu} H)(\overline{e_R}\gamma^{\mu}e_R)$ that gives rise to CLFV Higgs decays, because its coefficient is suppressed by a double chirality flip $\propto y_{\tau}^2$. 

All other one-loop induced operators are flavor conserving. Here, we do not discuss the corresponding observables in detail, because the resulting constraints are in general weaker than those from CLFV processes. In type-II seesaw, almost all flavor conserving operators are suppressed by $\mu_\Delta$, loop factors, and/or chirality flipping terms~\cite{Li:2022ipc}, so they are phenomenologically irrelevant. The only exception is provided by the lepton flavor conserving but LFU violating four-fermion operators in Eq.~\eqref{eftTree}, such as $(\overline{\mu}\gamma^{\mu}P_L \tau)(\overline{\tau}\gamma_{\mu}P_L \mu)$. LFU tests in $\tau$ decays constrain the  cut-off scale of such kind of operators to be above $4.5$~TeV~\cite{Allwicher:2023shc}. As we will see, this bound is still weaker than those we obtain from the CLFV $\tau$ decays. 


\section{Texture zeros and bounds on the type~II seesaw scale}
\label{texturesection}

In the charged-lepton mass basis, all six possible one-zero textures for the neutrino mass matrix $M^{\nu}$ are consistent with the neutrino mixing angles and mass splittings measured in neutrino oscillation experiments~\cite{Xing:2003ic, Xing:2004ik, Merle:2006du, Lashin:2011dn, Deepthi:2011sk, Bora:2016ygl, Kitabayashi:2020ajn, Chauhan:2023faf}: 
\begin{equation}
\label{onezero}
\begin{aligned}
&~ \mathcal{A^{}}: \left( \begin{matrix} 0 & b & a \cr b & y & c \cr a & c & z \end{matrix}\right) \; , \quad \mathcal{B^{}}: \left( \begin{matrix} x & 0 & a \cr 0 & y & c \cr a & c & z \end{matrix}\right) \; , \quad \mathcal{C^{}}: \left( \begin{matrix} x & b & 0 \cr b & y & c \cr 0 & c & z \end{matrix}\right) \;  ,\\
&~  \mathcal{D^{}}: \left( \begin{matrix} x & b & a \cr b & 0 & c \cr a & c & z \end{matrix}\right) \; , \quad { \mathcal{E}^{}}: \left( \begin{matrix} x & b & a \cr b & y & 0 \cr a & 0 & z \end{matrix}\right) \; , \quad { \mathcal{F}^{}}: \left( \begin{matrix} x & b & a \cr b & y & c \cr a & c & 0 \end{matrix}\right) \; .
\end{aligned}
\end{equation}
This is not the case for two-zero textures. The only patterns that, at present, are compatible with neutrino oscillation data are~\cite{Zhou:2015qua, Alcaide:2018vni,Singh:2019baq, Denton:2023hkx, Chauhan:2023faf,Treesukrat:2025dhd}:
\begin{equation}
\label{twozero}
\begin{aligned}
&~ \mathbf{A_1}: \left( \begin{matrix} 0 & 0 & a \cr 0 & y & c \cr a & c & z \end{matrix}\right) \; , \quad \mathbf{A_2}: \left( \begin{matrix} 0 & b & 0 \cr b & y & c \cr 0 & c & z \end{matrix}\right) \; , \quad \mathbf{B_1}: \left( \begin{matrix} x & b & 0 \cr b & 0 & c \cr 0 & c & z \end{matrix}\right) \; , \quad \mathbf{B_2}: \left( \begin{matrix} x & 0 & a \cr 0 & y & c \cr a & c & 0 \end{matrix}\right) \; ,\\
&~ \mathbf{B_3}: \left( \begin{matrix} x & 0 & a \cr 0 & 0 & c \cr a & c & z \end{matrix}\right) \; , \quad \mathbf{B_4}: \left( \begin{matrix} x & b & 0 \cr b & y & c \cr 0 & c & 0 \end{matrix}\right) \; , \quad \mathbf{C^{~}_{~}}: \left( \begin{matrix} x & b & a \cr b & 0 & c \cr a & c & 0 \end{matrix}\right) \; .
\end{aligned}
\end{equation}
Textures with three or more zeros are excluded by oscillation experiments~\cite{Xing:2004ik, Fritzsch:2011qv}. The above two-zero textures are also consistent with
beta decay and neutrinoless double beta decay ($0\nu\beta\beta$) data~\cite{Treesukrat:2025dhd}.\footnote{Only textures $\mathbf{A_1}$ and $\mathbf{A_2}$ are compatible with the stringent cosmological bound on the sum of neutrino masses $\Sigma m_{\nu} < 0.09$ eV~\cite{Denton:2023hkx}, while all seven textures remain viable under more conservative cosmological limits~\cite{Treesukrat:2025dhd}. We include all textures because the information from cosmology is indirect and can be modified by new physics irrelevant to the type~II seesaw. For instance, light dark sector particles can suppress the neutrino abundance and thereby relax the cosmological constraints, see~\revised{\cite{Farzan:2015pca, Benso:2024qrg, Das:2025asx}} and references therein for further examples.} They serve as the maximal possible hierarchal structure for $M^{\nu}$ and, in the context of type~II seesaw, $Y_{\Delta}$. 

In general, $M^{\nu}$ contains 9 real physical parameters, as shown by Eqs.~\eqref{eq:mnu} and \eqref{eq:pmns}: two mass differences $\Delta m_{12}$, $\Delta m_{13}$, three mixing angles $\theta_{12}$, $\theta_{13}$, $\theta_{23}$, one Dirac CPV phase $\delta$, two Majorana CPV phases $\rho$ and $\sigma$, and the lightest mass $m_{0}$, which is equal to $m_{\nu_1}$ in case of normal ordering (NO) or $m_{\nu_3}$ in case of inverted ordering (IO). Oscillation experiments can precisely measure the first five parameters, while they constrain the Dirac CPV phase $\delta$ only poorly at present and they are not sensitive to the absolute mass and the Majorana phases. Cosmological observations~\cite{Planck:2018vyg,DESI:2025zgx} and $\beta$-decay experiments such as KATRIN~\cite{KATRIN:2024cdt} can set upper limits to the absolute neutrino mass scale $m_{0}$.

In presence of a two-zero texture, $M^{\nu}$ only contains five independent real physical parameters, which we can choose to be the precisely measured ones: the mixing angles and the mass splittings \{$\theta_{12}$, $\theta_{13}$, $\theta_{23}$, $\Delta m_{12}$, $\Delta m_{13}$\}. The other physical parameters are predictions of the texture. The explicit relations among the entries of $M^{\nu}$ and the physical observables are shown in Appendix~\ref{expressions}. 
It is worth remarking that flavor patterns less restrictive than two-zero textures are only poorly constrained by data at present. Although $M^{\nu}$ with certain one-zero textures contains the same number of free parameters and experimental inputs, the non-zero elements can only be inferred with large uncertainties, unless $\delta$ and $m_0$ are measured or constrained at a better precision level. Typically, in order to narrow down the allowed ranges for the parameters in Eq.~\eqref{onezero}, the sensitivity for $m_0$ should reach the $\mathcal{O}(0.05)$~eV level so that the uncertainty of the dimensionless variable $(m_0^2/\Delta m_{13}^2)$ becomes smaller than $\mathcal{O}(1)$,
a result that is supported by the $m_0$ dependence of the $M^{\nu}$ elements~\cite{FileviezPerez:2008jbu}.

In the context of type~II seesaw, the two-zero textures $\mathbf{B_1}$, $\mathbf{B_4}$, and $\mathbf{C}$ all predict the $\mu\to \bar e ee$ decay to occur at the tree level, as $Y_{\Delta\,e\mu}$ and $Y_{\Delta\,ee}$ are both non-vanishing --- cf.~Eq.~\eqref{twozero} and Table~\ref{3body}. 
For all the other two-zero textures, $\mu\to \bar e ee$ vanishes at the tree level such that its BR is suppressed by a loop factor $1/(4\pi)^4=4\times10^{-5}$. 
The best probe for type~II seesaw then becomes $\mu\rightarrow e\gamma$, whose branching ratio is proportional to $(Y_{\Delta}Y_{\Delta}^{\dagger})_{\mu e}$ --- cf.~Table~\ref{3body} --- and is thus connected to the neutrino oscillation parameters by the following expression~\cite{Ardu:2023yyw}:
\begin{equation}
\label{universal}
    \text{BR}(\mu\to e\gamma)~\propto~\left|(Y_{\Delta}Y_{\Delta}^{\dagger})_{\mu e}\right|^2~\propto~ \left|(M^{\nu}M^{\nu\,\dagger})_{\mu e}\right|^2~\approx~ \left(\Delta m_{13}^2 \cos{\theta_{13}}\sin{\theta_{13}}\sin{\theta_{23}} \right)^2\,.
\end{equation}
This relation is valid for both normal and inverted ordering in the limit $\Delta m_{12}^2 \to 0$. Therefore, $\text{BR}(\mu\to e\gamma)$ only weakly depends on the unknown neutrino mass scale and Majorana phases, and is non-vanishing for all textures. 
As a result, simultaneously observing $\mu\to \bar e ee$, $\mu\to e\gamma$, and, in some cases, tree-level $\tau$ CLFV decays is, in principle, possible at current and upcoming experiments, as discussed below. This interesting situation is in stark contrast to the standard expectation for type~II seesaw with a generic flavor structure of $Y_\Delta$, where the strong limit set by $\mu\to \bar eee$ typically prevents the rates of other CLFV modes (especially for $\tau$ leptons) from resulting in the observable range; see~e.g.~\cite{Abada:2007ux,Dinh:2012bp} for detailed discussions based on the generic flavor structure.

\begin{table}[t!]
\renewcommand\arraystretch{2.}
    \centering
$\begin{array}{l|ccccccc}
\hline
 ~& \bf{A_1}& \bf{A_2}& \bf{B_1} & \bf{B_2}& \bf{B_3}& \bf{B_4} & \bf{C} \\
 \hline
 \mu \to \bar{e}ee & 8.3^{+ 0.6}_{-0.5} & 10.1^{+ 0.8}_{-0.9} & \textcolor{red}{22^{+ 7}_{-9}} & 3.3^{+ 0.7}_{-0.8} & 3.1^{+ 0.5}_{-0.6} & \textcolor{red}{20^{+ 5}_{-9}} &
   \textcolor{red}{75^{+ 49}_{-58}} \\
 \mu \to e\gamma  & \textcolor{red}{28.0^{+ 2.3}_{-1.9}} & \textcolor{red}{29.4^{+ 2.9}_{-4.0}} & 9.3^{+ 2.8}_{-4.0} & \textcolor{red}{9.8^{+ 3.5}_{-5.0}} & \textcolor{red}{8.9^{+ 2.6}_{-4.0}} & 9.5^{+ 3.3}_{-5.0} & 38^{+ 18}_{-35} \\
 \tau \to \bar{\mu }\mu \mu  & 6.1^{+ 0.7}_{-0.7} & 6.0^{+ 0.4}_{-0.4} & 0.5^{+0.2}_{-0.2} & 2.9^{+ 0.9}_{-1.3} & 0.2^{+0.2}_{-0.2} & 2.7^{+ 0.8}_{-1.3} & 0.4^{+0.5}_{-0.3} \\
 \tau \to \bar{\mu }ee & 0.64^{+0.04}_{-0.06} & 0.63^{+0.05}_{-0.06} & 5.7^{+ 0.2}_{-0.2} & 5.7^{+ 0.2}_{-0.2} & 5.7^{+ 0.2}_{-0.2} & 5.7^{+ 0.2}_{-0.2} &
   6.4^{+ 1.5}_{-1.7} \\
   \hline
\end{array} $
    \caption{Central values and $3\sigma$ uncertainties for the lower limit on the type~II seesaw scale $\Lambda_{\Delta}\equiv m_{\Delta}/(2|Y_{\Delta\mu\tau}|)$ (in TeV) set by various CLFV processes for the two-zero textures defined in Eq.~\eqref{twozero}. The uncertainties originate from the neutrino oscillation parameters. \revised{The neutrino mass ordering (NO/IO) is a prediction of each texture, as discussed in Appendix~\ref{expressions}.} For each texture, the strongest constraint is marked in red. }
    \label{cutoff}
\end{table}

Since $Y_{\Delta\mu\tau}$ is non-vanishing across all seven allowed two-zero textures, we define the type~II seesaw scale as\footnote{\rerevised{Notice that the scale $\Lambda_{\Delta}$ that we define here is the cut-off scale of the dimension-six operators relevant for CLFV, while for what concerns the renormalization group evolution of the parameters, it is the running from $\Lambda_{\text{UV}}$ to $m_{\Delta}$ that matters, as discussed in Section~\ref{RGstability}.}}:
\begin{equation}
\label{cut}
    \Lambda_{\Delta}~\equiv~\frac{m_{\Delta}}{2|Y_{\Delta \mu\tau}|}\,,
\end{equation}
where the factor 2 accounts for the fact that $Y_{\Delta\mu\tau}=Y_{\Delta\tau\mu}$ and both entries lead to the same interaction term. Other non-zero Yukawa couplings have values of the same order of magnitude as $Y_{\Delta \mu\tau}$, and can be connected to $Y_{\Delta \mu\tau}$ by neutrino oscillation observables, as illustrated in Appendix~\ref{expressions}. 
The $\mu\to \bar e ee$ constraint requires $\Lambda_{\Delta}$ larger than $\mathcal{O}(100)$ TeV~\cite{Abada:2007ux,Dinh:2012bp}, if all entries $Y_{\Delta}$ are of the same order of magnitude, that is, for a democratic (or anarchical) flavor structure. As mentioned above, this is no longer the case for some of the two-zero textures. 

In order to compare the predictions of the seven textures in Eq.~\eqref{twozero}, we calculate the current lower bounds on $\Lambda_{\Delta}$ for each texture as follows.
First, we fix $m_{\Delta}=3$ TeV since varying $m_{\Delta}$ within the range $1-10$~TeV does not significantly impact on the numerical value of Eq.~\eqref{loopf}.
Then, for each texture, we calculate the relation between $Y_{\Delta\mu\tau}$ and the other entries of $Y_{\Delta}$ using the neutrino oscillation data.
Due to the experimental uncertainties of the latter, we generate $10^4$ neutrino oscillation data sets that follow the distribution fitted by Nufit-6.0~\cite{Esteban:2024eli} and use them to calculate $10^4$ benchmarks for ${Y_{\Delta ij}}/{Y_{\Delta\mu\tau}}$ --- we refer to Appendix~\ref{expressions} for more details, \revised{including the neutrino mass ordering predicted in each texture.}
Using the expressions shown in Table~\ref{3body}, these benchmark values can be used to obtain the various CLFV BRs as a function of $Y_{\Delta\mu\tau}$ only. Here, one-loop expressions are applied only when the tree-level contribution vanishes. 
Next, we use the experimental limits collected in the last column of Table~\ref{3body} to constrain each $Y_{\Delta\mu\tau}$ benchmark and translate them to the lower limits for $\Lambda_{\Delta}$.
Finally, we get $10^4$ possible lower bounds on $\Lambda_\Delta$ as output for each texture and process, and show the corresponding central values in Table~\ref{cutoff}.
The uncertainty displayed for each entry of the table represents the range that 99.7~\% of the output numbers lie within ($3\sigma$ range). 
As we can see, the variance for texture $\mathbf{C}$ is especially sizable because, within this texture, the entry $Y_{\Delta\mu e}$ can be tuned to be small when the CPV phase $\delta$ takes specific values. 

In Table~\ref{cutoff}, we highlight in red the strongest bound for each texture, in all cases coming from $\mu\to e$ transitions. 
These results inform us that the textures $\mathbf{A_1}$, $\mathbf{A_2}$, $\mathbf{B_1}$, $\mathbf{B_4}$, and $\mathbf{C}$ are inconsistent with an effective scale $\Lambda_{\Delta}\lesssim 10$~TeV, which, of course, does not imply that $m_\Delta$ cannot be in the TeV range if $Y_{\Delta\mu\tau}$ is sufficiently small.
On the other hand, $\mathbf{B_2}$ and $\mathbf{B_3}$ are more effective in suppressing $\mu\to e$ transitions, so that they allow $m_\Delta$ to be as light as of $5-6$~TeV even for $Y_{\Delta\mu\tau}\approx 0.5$.\footnote{This is to be compared to the direct LHC limit $m_\Delta \gtrsim 1$~TeV --- see Section~\ref{sec:collider}.}  
In addition, the table shows that fulfilling the $\mu\to \bar e ee$ constraint does not always require $\Lambda_{\Delta}\gtrsim \mathcal{O}(100)$~TeV as expected for a democratic flavor structure, no matter whether this process is induced at tree or one-loop level. Table~\ref{cutoff} also shows that $\mu\to e\gamma$ provides tight constraints for all the textures, but still up to large uncertainties. Meanwhile, combined $\tau$ CLFV decays provide a universal bound $\Lambda_{\Delta}\gtrsim 5-6$~TeV for all textures. As we can see for $\mathbf{B_2}$ and $\mathbf{B_3}$, the scales probed by $\tau \to\bar \mu ee$ and $\mu\to e\gamma$ are comparable within uncertainties.

We conclude this section with a brief comparison between the results presented above and some well-studied frameworks introducing flavor symmetries in the leptonic sector.
A widely used framework to suppress 2-1 flavor changing processes relative to 3-2 and 3-1 transitions involves the assumption that the SMEFT operators (thus the NP interactions) are invariant under the same $U(2)^5$ global symmetry approximately displayed by SM flavor sector, which acts on the lightest two generations only. 
\revised{See Ref.~\cite{Covone:2025lee} for a recent analysis on the charged lepton sector.} 
If unbroken, this symmetry suppresses all $\mu\to e$ transitions because of $\mu-e$ flavor universality. 
In addition, the $U(2)^5$ symmetry is chiral so that dipole operators do not involve the two lightest lepton families. In this framework, the strongest bound for the cutoff scale in the lepton sector is about 5~TeV, mainly following from LFU tests in $\tau$ decays~\cite{Allwicher:2023shc, Allwicher:2025bub}.
This bound can be further relaxed to values even closer to the TeV scale if the NP couplings to $e$ and $\mu$ are mildly suppressed~\cite{Allwicher:2023shc}. 
Here we want to highlight that, for textures $\mathbf{B_2}$ and $\mathbf{B_3}$,  constraints on the NP scale comparable to those within the $U(2)^5$ framework are obtained, even if $\Delta$ couples to $e$ and $\mu$ non-universally. 
Textures $\mathbf{B_2}$ and $\mathbf{B_3}$ are simple examples showing that TeV-scale NP can be consistent with the $\mu\to e$ constraints, even without flavor symmetries. 

Next, let us remark that $\mathbf{B}$ textures feature a moderate hierarchy, as one can also see from the numerical benchmark in Eq.~\eqref{twozerobenchmark}. As lowest order approximation, the normalized Yukawa matrix can, in fact, be written as:
\begin{equation}
\label{A40}
\frac{|Y_{\Delta ij}|}{|Y_{\Delta\mu\tau}|}~\approx~\left(
    \begin{matrix}
       1 & 0 & 0\\
        0 & 0 & 1 \\
        0 & 1 & 0\\
    \end{matrix}\right), \quad \text{for textures}~\bf{B_1}-\bf{B_4}\,.
\end{equation}
This structure can originate from flavor symmetries such \revised{as} $A_4$ and its subgroup $Z_3$~\cite{Dev:2014dla, Bigaran:2022giz, Lichtenstein:2023iut, Lichtenstein:2023vza, Palavric:2024gvu, Moreno-Sanchez:2025bzz}. For instance, the finite transformation:
\begin{equation}
    \label{Z3A4}
    \Delta\to \Delta, \quad e\to e, \quad \mu \to e^{i\frac{2\pi}{3}}\mu, \quad  \tau\to e^{i\frac{4\pi}{3}}\tau, 
\end{equation}
can protect the stability of the zero texture shown in Eq.~\eqref{A40} without extending our model beyond the minimal type~II seesaw, and ensures that all $\mu\to e$ transitions vanish. The only CLFV operator arising at the tree level is $(\overline{\tau}\gamma^{\mu}P_L e)(\overline{\mu}\gamma_{\mu}P_L e)$, whose direct bound comes from a third-generation related process, the $\tau\rightarrow \overline{\mu} ee$ decay. On the other hand, we notice that the $A_4$ (or $Z_3$) discrete symmetry is also not strictly necessary in order to relax the lower limit of $\Lambda_{\Delta}$ down to the TeV scale. The benchmark textures shown in Eq.~\eqref{twozerobenchmark} demonstrate that $Y_{\Delta\mu\mu}$ or $Y_{\Delta\tau\tau}$ can be as large as $0.3-0.4$. These entries serve as sizable $A_4$ or $Z_3$ breaking terms while not leading to overwhelming constraints from $\mu\to e$ transitions.


\section{Radiative stability of texture zeros}
\label{RGstability}

It is known that $M^{\nu}$ texture zeros are absolutely stable below $m_{\Delta}$~\cite{Hagedorn:2004ba, Fritzsch:2011qv}. Indeed, the relevant RG equation describing the evolution of the matrix $M^{\nu}$ with the scale $\mu$ reads~\cite{Chankowski:1993tx, Babu:1993qv}:
\begin{equation}
\label{MRGE}
\begin{aligned}
    16\pi^2\frac{d M^{\nu}}{d\log{\mu}}~=&~-\frac{1}{2}\left( M^{\nu}Y_{\ell}Y_{\ell}^{\dagger}+(Y_{\ell}Y_{\ell}^{\dagger})^T M^{\nu}\right)+ M^{\nu}\left(-3 g_2^2+2 \lambda_H+6y_t^2 \right)\,, 
\end{aligned}
\end{equation}
where $g_2$, $\lambda_H$, and $y_t$ are respectively the $SU(2)_L$ gauge coupling, the Higgs quartic coupling, and the top-quark Yukawa coupling. $Y_{\ell}$ is the Yukawa matrix of charged leptons. Other Yukawa couplings are far smaller than $\mathcal{O}(1)$ and cannot change the flavor structure of $M^\nu$, so they can be neglected to very good approximation. Working in the basis where $Y_{\ell}=\text{diag}\{y_e, y_{\mu}, y_{\tau}\}$, Eq.~(\ref{MRGE}) can be solved analytically~\cite{Balaji:2000ma}:
\begin{equation}
\label{sol0}
    M^{\nu}(\mu)~=~I_{g \lambda t} 
    \left( \begin{matrix}I_e^2  M^{\nu}_{ee}(m_{\Delta}) &I_e I_{\mu} M^{\nu}_{e\mu}(m_{\Delta}) &I_e I_{\tau} M^{\nu}_{\tau e}(m_{\Delta}) \cr I_{\mu}I_e M^{\nu}_{\mu e}(m_{\Delta}) & I_{\mu}^2 M^{\nu}_{\mu\mu}(m_{\Delta}) & I_{\mu} I_{\tau} M^{\nu}_{\tau\mu}(m_{\Delta}) \cr  I_{\tau} I_{e} M^{\nu}_{\tau e}(m_{\Delta}) & I_{\tau} I_{\mu} M^{\nu}_{\tau \mu}(m_{\Delta}) & I_{\tau}^2 M^{\nu}_{\tau \tau}(m_{\Delta}) \end{matrix} \right).
\end{equation}
The renormalization factors $I_{g\lambda t}(\mu)$ and $I_\ell(\mu)$, whose expressions are shown in Appendix~\ref{overall}, are equal to one for $\mu=m_{\Delta}$. This solution implies that radiative corrections to $M^{\nu}$ are always multiplicative: as long as $M^{\nu}_{\ell \ell'}$ is zero at one scale below $m_{\Delta}$, it vanishes at all scales below $m_{\Delta}$. 

\begin{figure}[t!]
  \centering
  \includegraphics[width=0.7\textwidth]{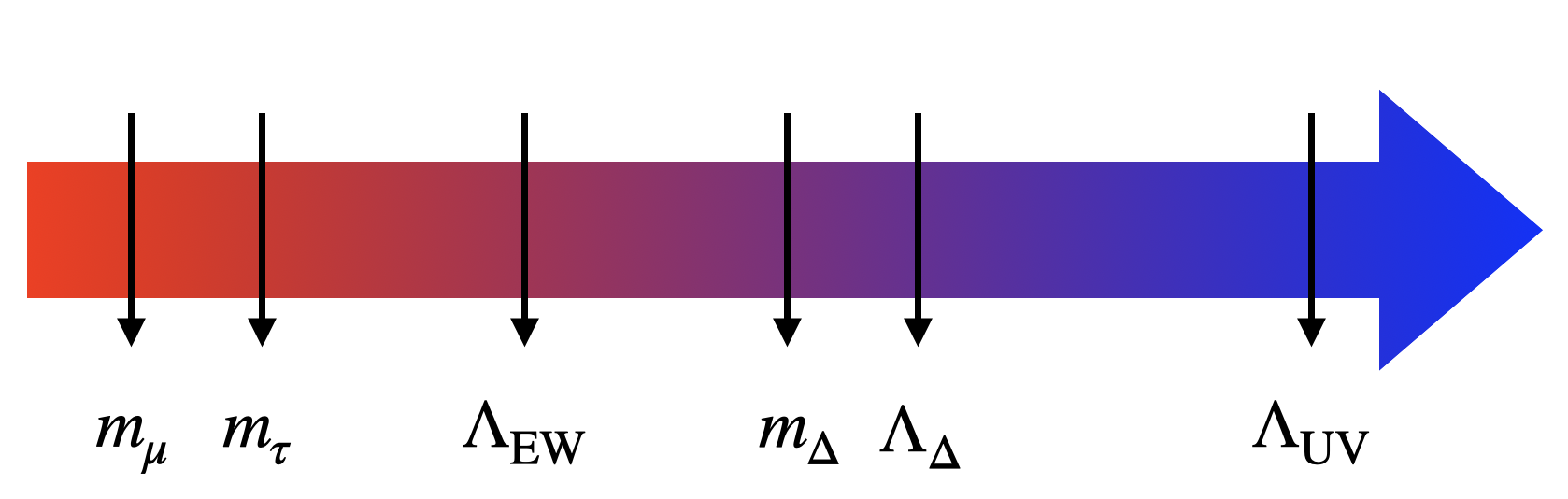}
    \caption{Schematic (approximately logarithmic) representation of the energy scale hierarchy in the considered type~II seesaw framework. Vertical arrows indicate the key energy scales: muon and tau masses, $m_\mu$ and $m_\tau$, the  electroweak symmetry breaking scale $\Lambda_\text{EW}$, the mass of the type~II seesaw scalar triplet $m_{\Delta}$, \rerevised{the cut-off scale $\Lambda_{\Delta}$ of the dimension-six operators relevant for CLFV}, and the ultraviolet cutoff scale $\Lambda_\text{UV}$.}
   \label{scales}
\end{figure}

On the other hand, texture zeros are typically not stable under the RG running above $m_{\Delta}$. In other words, the considered textures can be consistently defined only at the high-energy scale $\Lambda_\text{UV}$ where they originate (e.g.~from spontaneous breaking of a flavor symmetry) and, unless $\Lambda_\text{UV} = m_\Delta$, radiative effects will, in general, generate non-zero entries.
The scales relevant to the problem are illustrated in Figure~\ref{scales}.
The relevant RG equations above the scale $m_\Delta$ are~\cite{Chao:2006ye, Schmidt:2007nq}:
\begin{equation}
\label{rges}
    \begin{aligned}
        16\pi^2\frac{d Y_{\ell}}{d\log{\mu}}~=&~ 3Y_{\ell}Y_{\Delta}^{\dagger}Y_{\Delta}+Y_{\ell}\left(3y_t^2-\frac{9}{4}g_1^2-\frac{9}{4}g_2^2\right)\,,\\
        16\pi^2\frac{d Y_{\Delta}}{d\log{\mu}}~=&~6Y_{\Delta}Y_{\Delta}^{\dagger}Y_{\Delta} 
        +Y_{\Delta}\left(2\text{tr}\left[Y_{\Delta}^{\dagger}Y_{\Delta}\right]-\frac{9}{10}g_1^2-\frac{9}{2}g_2^2\right)\,,
    \end{aligned}
\end{equation}
where, as before, we neglect all the couplings much smaller than $\mathcal{O}(1)$, including $Y_{\ell}^{\dagger}Y_{\ell}\sim y_{\tau}^2$. 
As mentioned, if a given zero texture for $Y_{\Delta}$ originates at some scale $\Lambda_\text{UV} > m_{\Delta}$, running effects from $\Lambda_\text{UV}$ down to $m_{\Delta}$ can change the $Y_\Delta$ flavor structure. The resulting matrices at $m_{\Delta}$ are:
\begin{equation}
\label{sol1}
\begin{aligned}
    Y_{\ell}(m_{\Delta})~=&~I_{tg}  Y_{\ell}(\Lambda_\text{UV})\cdot \left(\frac{m_{\Delta}}{\Lambda_\text{UV}}\right)^{\frac{3}{16\pi^2}\left(Y_{\Delta}^{\dagger}Y_{\Delta}\right)_{\text{eff}}}\,, \\
    Y_{\Delta}(m_{\Delta})~=& ~I_{\Delta g}\left(\frac{m_{\Delta}}{\Lambda_\text{UV}}\right)^{\frac{3}{16\pi^2}\left(Y_{\Delta}^{\dagger}Y_{\Delta}\right)^T_{\text{eff}}}\cdot  Y_{\Delta}(\Lambda_\text{UV})\cdot \left(\frac{m_{\Delta}}{\Lambda_\text{UV}}\right)^{\frac{3}{16\pi^2}\left(Y_{\Delta}^{\dagger}Y_{\Delta}\right)_{\text{eff}}}\,.
\end{aligned}
\end{equation}
The notation used here is explained in detail in Appendix~\ref{overall}. 
$I_{tg}$ and $I_{\Delta g}$ are overall factors accounting for the flavor-conserving contribution and are given in Eq.~\eqref{Idef}. The effective matrix $\left(Y_{\Delta}^{\dagger}Y_{\Delta}\right)_{\text{eff}}$, defined in Eq.~(\ref{Ydef}), can be interpreted as the average of $\left(Y_{\Delta}^{\dagger}Y_{\Delta}\right)$ between the scales $m_{\Delta}$ and $\Lambda_\text{UV}$. 
The expressions in Eq.~\eqref{sol1} involve matrix functions, which can be effectively evaluated by matrix power expansion, as explained in and below Eq.~(\ref{matrixpower}). 

A new feature of the above solution is that $Y_{\ell}(m_{\Delta})$ in Eq.~\eqref{sol1} is no longer diagonal. To quantify how radiative corrections destabilize the zero textures, one then has to diagonalize $Y_{\ell}$ by rotating the $\ell_{L}$ and $e_{R}$ basis, which further modifies $Y_{\Delta}(m_{\Delta})$. Assuming $Y_{\Delta}$ to vary slowly with respect to $\mu$, we derive the leading-log expansion for $Y_{\Delta}(m_{\Delta})$ in the basis of diagonal $Y_{\ell}$:
\begin{equation}
\label{Ydiag}
    Y_{\Delta}(m_{\Delta})=I_{\Delta g} \left(Y_{\Delta}(\Lambda_\text{UV}) +\frac{3}{16\pi^2}\log\left(\frac{m_{\Delta}}{\Lambda_\text{UV}}\right) \left(2Y_{\Delta} Y_{\Delta}^{\dagger}Y_{\Delta}-(Y_{\Delta}^{\dagger}Y_{\Delta})_{ij}\left[Y_{\Delta}, \mathcal{O}_{\{ij\}}\right]  \right)(\Lambda_\text{UV}) \right)\,. 
\end{equation}
To obtain this expression, we take into account the hierarchy $m_{\tau}\gg m_{\mu}\gg m_e$, which implies that $Y_{\ell}(m_{\Delta})$ in Eq.~\eqref{sol1} can be diagonalized by rotating the $\ell_{L}$ basis only. Here, $\mathcal{O}_{\{ij\}}$ are matrices defined in terms of the Gell-Mann matrices~\cite{Gell-Mann:1962yej} as $\mathcal{O}_{\{12\}}=i\lambda_2$, $\mathcal{O}_{\{13\}}=i\lambda_5$, $\mathcal{O}_{\{23\}}=i\lambda_7$, while the others are zero. 

We proceed assuming that $Y_{\Delta}(\Lambda_\text{UV})$ displays one of the zero textures in Eq.~(\ref{onezero}) and Eq.~(\ref{twozero}), plug it into the solution Eq.~(\ref{Ydiag}), and check whether such texture is preserved for $Y_{\Delta}(m_{\Delta})$. We find that, starting with $Y_{\Delta ee}(\Lambda_\text{UV}) =Y_{\Delta\mu e}(\Lambda_\text{UV})=0$, one obtains \mbox{$Y_{\Delta ee}(m_{\Delta})=Y_{\Delta \mu e}(m_{\Delta})=0$}, hence the RG running does not give rise to the vanishing entries. In addition, if $Y_{\Delta \mu e}(\Lambda_\text{UV})\neq 0$ but $Y_{\Delta ee}(\Lambda_\text{UV}) =0$, the relation $Y_{\Delta ee}(m_{\Delta})=0$ still holds. 

We further checked these features numerically and approximately confirmed the analytical results discussed above. Taking $\left(Y_{\Delta}^{\dagger}Y_{\Delta}\right)_{\text{eff}}$ in Eq.~\eqref{sol1} as a random complex matrix whose elements vary between $-1$ and $1$ (for both real and imaginary parts), we find that, if $Y_{\Delta ee}(\Lambda_\text{UV})=Y_{\Delta\mu e}(\Lambda_\text{UV})_{}=0,~ Y_{\Delta \tau\mu}(\Lambda_\text{UV})=1$, the entries $Y_{\Delta}(m_{\Delta})_{ee}$ and $Y_{\Delta}(m_{\Delta})_{\mu e}$ take values of the order $(m_e/m_{\tau})^2\sim10^{-7}$ and $(m_{\mu}/m_{\tau})^2\sim10^{-3}$, respectively. As a consequence, we can conclude that the zero textures $\mathcal{A^{}}$ and $\mathbf{A_1}$ remain approximately stable for any scale below $\Lambda_\text{UV}$. 
Although we cannot identify a simple symmetry responsible for this result, we note that one-loop RG stability is not necessarily enforced by symmetries. In some cases, special flavor structures can also forbid loop corrections up to certain orders. Known examples include the rank of the neutrino mass matrix~\cite{Zhang:2024weq} and $CP$ violation in extended Higgs sectors~\cite{Fontes:2021znm, deLima:2024lfc}. On the other hand, the other zero textures we considered are not RG stable. The elements that are zero at $\Lambda_\text{UV}$ receive additive corrections at one loop. These radiative contributions must be included in the phenomenological analysis and, as we will see, introduce some dependence of the results on the cutoff scale $\Lambda_\text{UV}$. 

\section{Patterns of CLFV observables from two-zero textures}
\label{phenosection}

\begin{figure}[t!]
  \centering
  \includegraphics[width=1\textwidth]{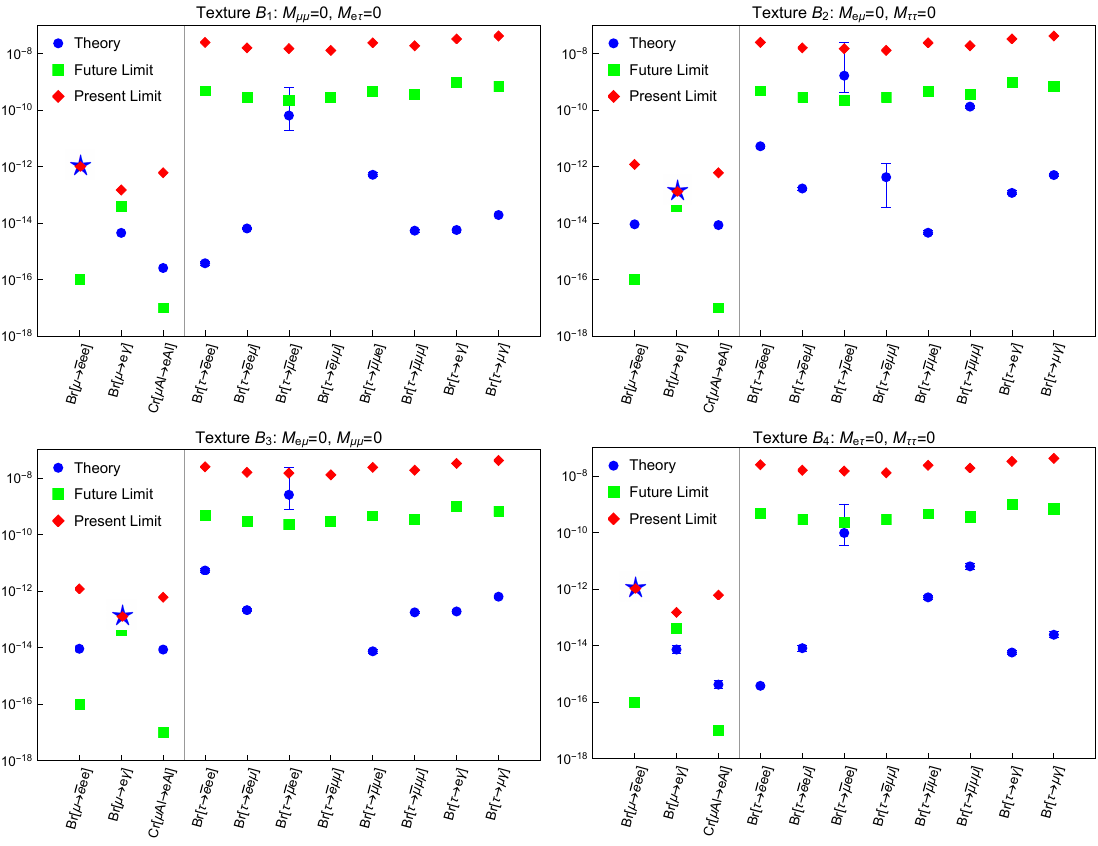}
      \caption{Predicted branching ratios of various CLFV processes for textures $\mathbf{B_1}$-$\mathbf{B_4}$. The strongest constraint, $\text{BR}(\mu\to e\gamma)$ or $\text{BR}(\mu\to \bar e ee)$, is fixed to the value that saturates the current experimental limit (blue star). The predicted BRs for the other processes are denoted by blue dots with error bars corresponding to $3\sigma$ uncertainties from neutrino oscillation data. Red diamonds and green squares respectively indicate current and future experimental sensitivities. See the main text for details. }
   \label{B1B4}
\end{figure}

\begin{figure}[t!]
  \centering
  \includegraphics[width=1\textwidth]{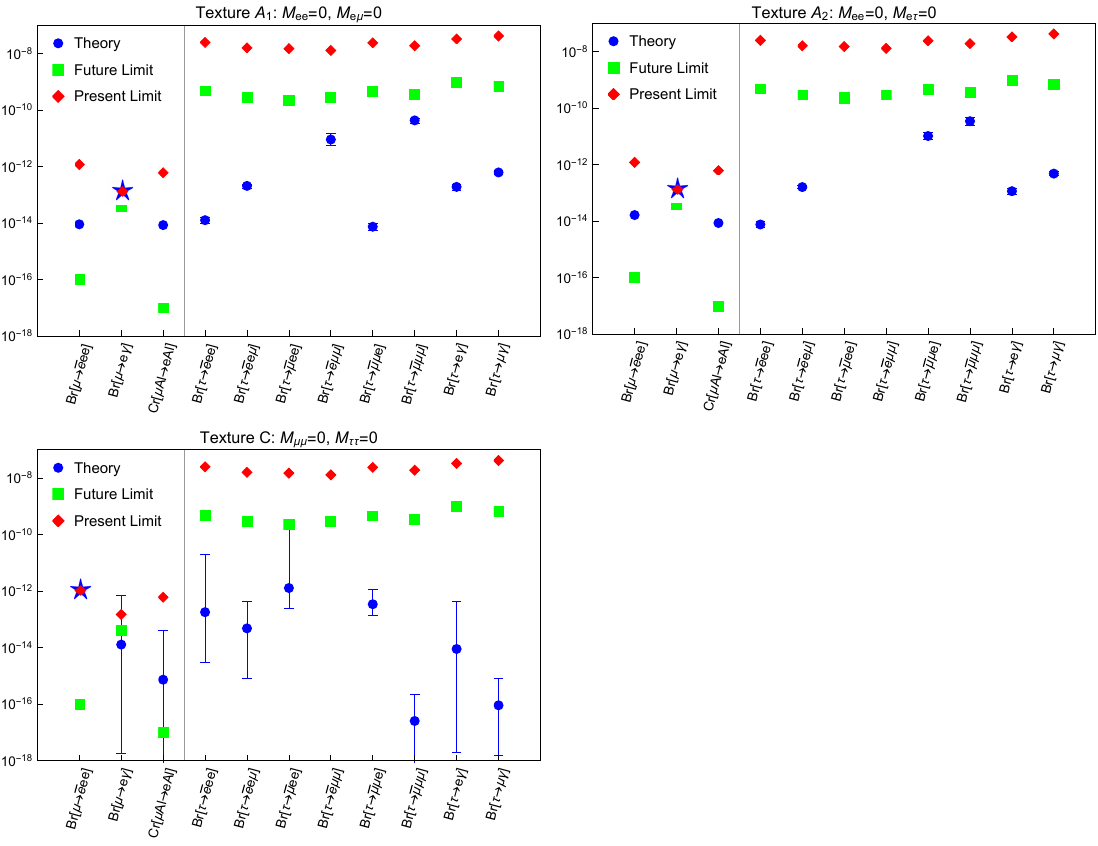}
    \caption{Same as Figure~\ref{B1B4} for textures \revised{$\mathbf{A_1}, \mathbf{A_2}$} and $\mathbf{C}$.}
   \label{A1A2C}
\end{figure}

In this section, we study in greater detail the phenomenology of type~II seesaw with $Y_\Delta$ featuring a two-zero texture, with the aim of illustrating the distinctive patterns that CLFV observables can display for each texture. 

Our main results are shown in \revised{Figures~\ref{B1B4} and~\ref{A1A2C}}, which illustrate the correlations among the relevant CLFV processes for the seven possible two-zero textures in Eq.~\eqref{twozero}, together with current and future experimental sensitivities. 
In these figures, we first set the observable providing the strongest constraint to its current experimental limit and mark it as a blue star. That is, we fix $\text{BR}(\mu\to \bar e ee)=1.2\times10^{-12}$ (the SINDRUM limit~\cite{SINDRUM:1987nra}) for textures $\mathbf{B_1}$, $\mathbf{B_4}$ and $\mathbf{C}$, and $\text{BR}(\mu\to e\gamma)=1.5\times10^{-13}$ (the recent MEG~II limit~\cite{MEGII:2025gzr}) for textures $\mathbf{A_1}$, $\mathbf{A_2}$, $\mathbf{B_2}$, $\mathbf{B_3}$.
In order to illustrate correlations with other processes \emph{independent} of the absolute size of $|Y_\Delta|$, we assume $|Y_{\Delta\mu\tau}|$ to be small enough that the terms proportional to $(Y_{\Delta}Y_{\Delta}^{\dagger})_{ee}$ contained in the expression for $\text{BR}(\mu\to \bar e ee)$ shown in Table~\ref{3body} can be neglected.\footnote{For $\mathbf{B_2}$, $\mathbf{B_3}$, this requires $|Y_{\Delta\mu\tau}|\ll 0.7$, which is the value for which those terms become dominant. This can be relaxed to about $1.5$ for \revised{$\mathbf{A_1}, \mathbf{A_2}$}.} 
This also ensures that the RG effects discussed in Section~\ref{RGstability} are so small that they do not destabilize the textures, as we will see in the following.
As a consequence of this assumption, all relevant CLFV BRs, no matter if they are dominated by tree-level or one-loop amplitudes, scale as $|Y_{\Delta\mu\tau}|^4$, and the ratios between different BRs are independent of the absolute size of the Yukawa couplings. 
Therefore, in this limit, all CLFV BRs within the same zero texture are connected by neutrino oscillation data only. The error bars illustrate the $3\sigma$ uncertainties arising from oscillation parameters, which we calculate using the same method as for the bounds on $\Lambda_{\Delta}$ discussed in Section~\ref{texturesection}. The red diamonds represent the current experimental limits for each process, as listed in Table~\ref{3body}.\footnote{For $\mu \to e$ conversion in nuclei, we use the expressions in Table~\ref{3body} to rescale the bound on  $\text{CR}(\mu\,\text{Au}\to e \,\text{Au})$ translating it into a limit on $\text{CR}(\mu\,\text{Al}\to e \,\text{Al})$.} Expected future sensitivities~\cite{MEGII:2018kmf,Blondel:2013ia,Kuno:2013mha,Mu2e:2014fns,Banerjee:2022xuw} are represented as green squares.

A few comments about the results shown in \revised{Figures~\ref{B1B4} and~\ref{A1A2C}} are in order. For textures $\mathbf{B_1}$, $\mathbf{B_4}$, $\mathbf{C}$, the $\mu\to \bar e ee$ decay arises at tree level and thus is the most sensitive probe of type~II seesaw. Observing $\mu\to e\gamma$ within the sensitivity of the MEG~II experiment~\cite{MEGII:2018kmf} would then directly exclude $\mathbf{B_1}$ and $\mathbf{B_4}$. On the other hand, texture $\mathbf{C}$ still allows for the simultaneous observation of $\mu\to \bar e ee$ and $\mu\to e\gamma$ due to the large theoretical uncertainty. 
For the remaining textures, the loop induced process $\mu\to e\gamma$ provides the strongest constraint on the model because it cannot be accidentally suppressed by our flavor structures. 
However, even if $\mu\to e\gamma$ is observed in the near future, the $\mu\to \bar e ee$ rate is predicted to be within the future experimental sensitivity, due to the significant projected improvement of the Mu3e experiment~\cite{Blondel:2013ia}. In addition, all textures predict that, as long as either $\mu\to e\gamma$ or $\mu\to \bar e ee$ are discovered, observing $\mu\to e$ conversion in nuclei would be ensured at the upcoming COMET~\cite{Kuno:2013mha} and Mu2e~\cite{Mu2e:2014fns} experiments.

In the $\tau$ sector, all textures of type $\mathbf B$ and (marginally) texture $\mathbf C$ allow the possibility of detecting $\tau\to\overline{\mu}ee$ at Belle~II~\cite{Banerjee:2022xuw,Belle-II:2018jsg}, even under the strict constraints from $\mu\to e$ transitions. This is consistent with general predictions based on $Z_3$ flavor symmetries, as discussed at the end of Section~\ref{texturesection}, while our two-zero textures are more general. In particular, textures $\mathbf{B_2}$ and $\mathbf{B_3}$ ensure detection of $\tau\to \overline{\mu}ee$ at Belle~II, given that $\mu\to e \gamma$ is observed. Current neutrino data even allows the possibility that $\tau\to \overline{\mu}ee$ be discovered before $\mu \to e\gamma$. 
In case either $\mu\to e \gamma$ or $\tau\to \overline{\mu}ee$ is discovered, both $\mu\to \bar e ee$ and $\mu \to e$ conversion would also be within the sensitivity of upcoming experiments.
Interestingly, although textures $\mathbf{A_1}$ and $\mathbf{A_2}$ can be interpreted as originated from a $U(1)_e$ symmetry with a (rather large) explicit breaking term --- see the benchmark in Eq.~\eqref{twozerobenchmark}\,--- the predicted rate of the most sizable $\tau$ CLFV  process, $\tau\to \bar\mu\mu\mu$, remains below the expected Belle~II sensitivity. In general, our figures show that observing this process (as well as any $\tau$ decay besides $\tau\to \overline{\mu}ee$) would exclude all the considered textures within type~II seesaw.
It is worth remarking that \revised{Figures~\ref{B1B4} and~\ref{A1A2C}} also imply that the less restrictive one-zero textures, such as $\mathcal A$ and $\mathcal B$ in Eq.~\eqref{onezero}, are not predictive within the present knowledge of neutrino parameters. For instance, one can see that textures $\mathbf{A_1}$ and $\mathbf{B_2}$ both satisfy the one-zero structure $\mathcal B$, but they give distinctive predictions for $\tau\to\overline{\mu}ee$.%

\begin{figure}[t!]
  \centering
  \includegraphics[width=0.98\textwidth]{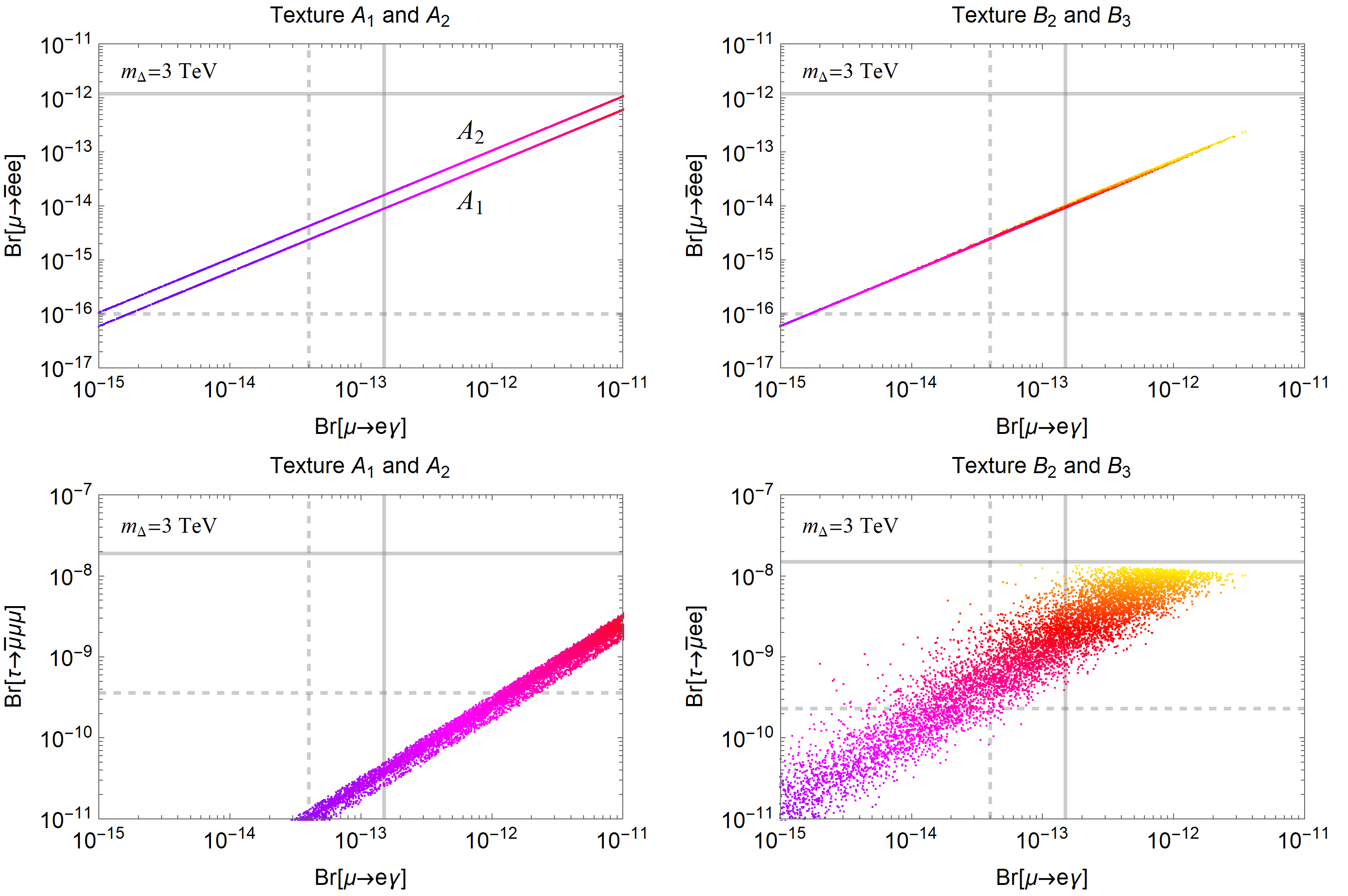}
 \includegraphics[width=0.4\textwidth]{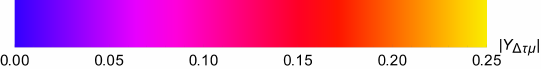}
    \caption{Predicted $\text{BR}(\mu\to e\gamma)$ within textures $\mathbf{A_1}$, $\mathbf{A_2}$, $\mathbf{B_2}$, $\mathbf{B_3}$, versus the second most constraining observable, $\text{BR}(\mu\to \bar e ee)$ (for all textures) and $\text{BR}(\tau \to\bar \mu ee)$ (for $\mathbf{B_2}$, $\mathbf{B_3}$ only). The parameters are set to $m_{\Delta}=3$~TeV and  $0\le |Y_{\Delta\tau\mu}| \le 0.25$. The other entries of $Y_\Delta$ are calculated as in Appendix~\ref{expressions}. Solid (dashed) lines represent current (future) experimental limits.
    }
   \label{scat}
\end{figure}

The results presented above show how each two-zero texture predicts a peculiar pattern for the CLFV processes within type~II seesaw, enabling to experimentally \revised{discriminate} between them in case several processes are observed.
This motivates us to study the absolute CLFV rates in more detail focusing on those textures that suppress the tree-level contribution to $\mu\to\bar eee$, that is, $\mathbf{A_1}$, $\mathbf{A_2}$, $\mathbf{B_2}$, and $\mathbf{B_3}$. 
The scatter plots in Figure~\ref{scat} show how $\text{BR}(\mu\to e \gamma)$, $\text{BR}(\mu\to \bar e ee)$, $\text{BR}(\tau\to\overline{\mu}ee)$, and $\text{BR}(\tau\to\overline{\mu}\mu\mu)$ are correlated and how they depend on the absolute size of $Y_{\Delta}$ for the benchmark triplet mass $m_{\Delta}=3$~TeV.
To generate these plots, we randomly vary $|Y_{\Delta\tau\mu}|$ between 0 and 0.25 (each point's color \revised{denotes} the value of $|Y_{\Delta\tau\mu}|$ as indicated by the bar below the plots), and calculate the other entries of $Y_{\Delta}$ using the neutrino oscillation data as done before. The branching ratios are then obtained using the (full) expressions in Table~\ref{3body}. Solid and dashed lines represent the current and expected future limits, respectively.


Figure~\ref{scat} shows that, currently, $\mu\to e\gamma$ is consistently the strongest constraint for all the four considered textures. 
The spread of the points corresponds to the error bars in \revised{Figures~\ref{B1B4} and~\ref{A1A2C}}. The sizable uncertainty of the $\mathbf{B_2}$ and $\mathbf{B_3}$ prediction for $\text{BR}(\mu\to e\gamma)$ allows $Y_{\Delta\tau\mu}$ with values as large as $0.25$. 
Hence, these textures are still consistent with a TeV-scale $\Delta$ and sizable Yukawa couplings, enriching the phenomenology of the minimal type~II seesaw, especially in view of present and future collider searches.
On the other hand, the $\mu\to e\gamma$ constraint implies a more stringent limit $|Y_{\Delta\mu\tau}|\lesssim 0.05$ for the $\mathbf A$ textures.
In all cases displayed in Figure~\ref{scat}, if $\text{BR}(\mu\to e\gamma)$ is within the future experimental sensitivity, so is $\text{BR}(\mu\to \bar e ee)$. 
This encouraging result also applies to $\text{BR}(\tau\to \overline{\mu}ee)$ for textures $\mathbf{B_2}$, $\mathbf{B_3}$. By contrast, this is not the case for $\mathbf{A_1}$, $\mathbf{A_2}$, in which $\tau\to \overline{\mu}ee$ vanishes at tree-level and is not enhanced by the leading-log term at one loop.  
$\mathbf{A_1}$ and $\mathbf{A_2}$ predict $\tau\to \bar\mu\mu\mu$ to be the $\tau$ CLFV process with the largest rate, albeit far below the expected sensitivity. These textures can be then excluded if any of the $\tau$ CLFV decays is detected at Belle~II. 

Figure~\ref{scat} also illustrates the correlations between different processes more transparently than in our previous discussion. As one can see, $\mu\to \bar e ee$ and $\mu\to e\gamma$ are in all cases linearly correlated to very good approximation. 
Texture $\mathbf{A_2}$ predicts $\text{BR}(\mu\to \bar e ee)/\text{BR}(\mu\to e\gamma)$ nearly twice as large as the value predicted by $\mathbf{A_1}$, $\mathbf{B_2}$, and $\mathbf{B_3}$, because it features $\mu$ instead of $\tau$ running in the penguin loop (diagram (g) of Figure~\ref{Feynmandiagrams}), which enhances the log term in the decay amplitude. 
For all considered textures, on the other hand, the correlation between $\mu$ and $\tau$ decays is less sharp because of the dependence on the neutrino oscillation data.


\begin{figure}[t!]
  \centering
  \includegraphics[width=1\textwidth]{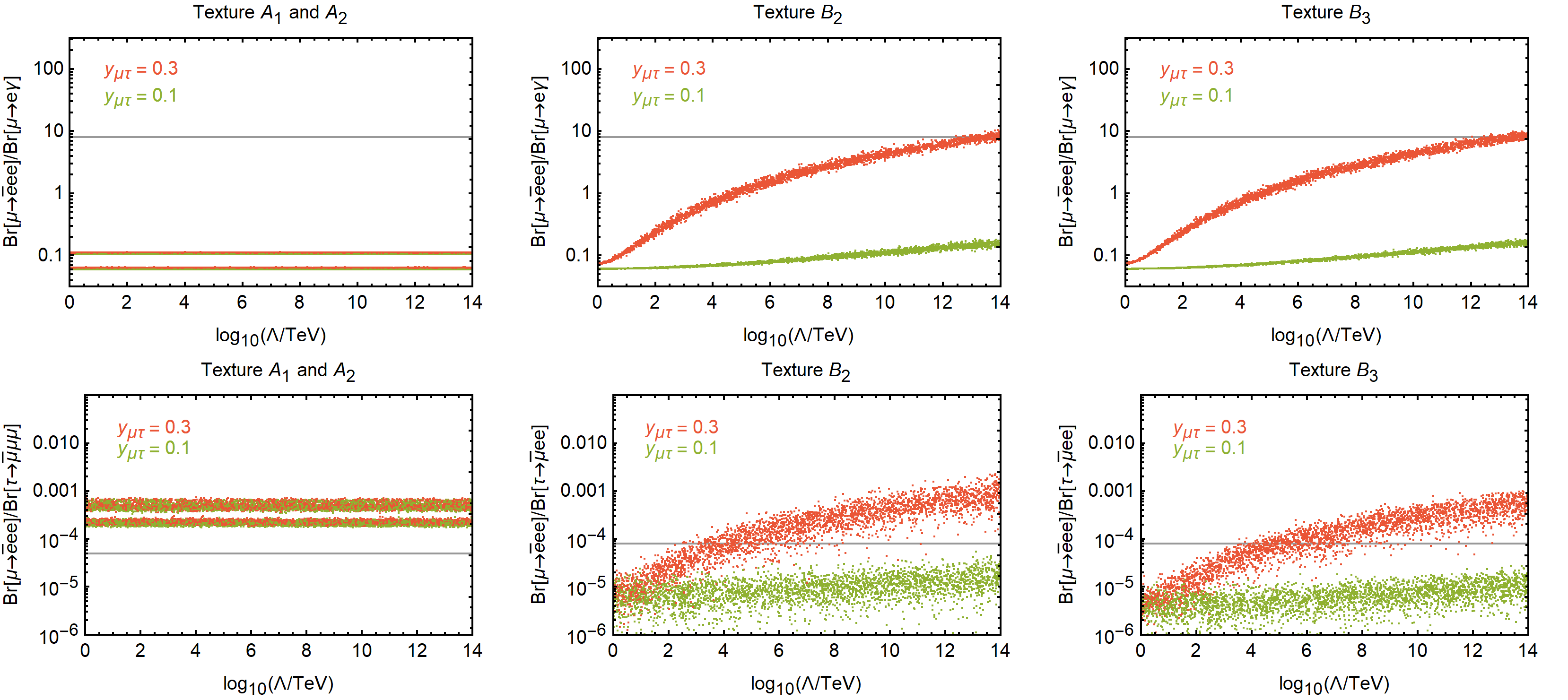}
    \caption{Ratios of CLFV BRs as functions of the scale $\Lambda_\text{UV}$ where a given texture is assumed to be exact, including the RG evolution of $Y_\Delta$ above $m_{\Delta}$. Green points correspond to $Y_{\Delta\mu\tau}=0.1$, red points to $Y_{\Delta\mu\tau}=0.3$. The gray solid lines show the ratios of the current experimental limits.}
   \label{RGE2}
\end{figure}

So far, we have not included in the analysis the RG corrections discussed in Section~\ref{RGstability}.
In general, the RG corrections to the `would-be' zero elements in $Y_{\Delta}$ are negligible for what concerns neutrino masses and mixing, so that the formulae in Appendix~\ref{expressions} remain valid and the non-zero elements are still correlated by neutrino oscillation data. As a consequence, the leading non-zero values of $Y_{\Delta}(m_{\Delta})$ can always be understood as experimental inputs and definitions.
However, the would-be zero elements receive additive RG corrections as discussed in Section~\ref{RGstability} and, for the physical observables depending on these elements, such corrections can become phenomenologically important even if small.

Figure~\ref{RGE2} shows ratios of the CLFV BRs including the RG effects, with the gray lines indicating the ratios of the present experimental limits. Hence, the process in the numerator (denominator) sets the stronger constraint for a point lying above (below) such line.
For each point in the plots, we choose $Y_{\Delta\mu\tau}(m_{\Delta})=0.1$ (green) and 0.3 (red) as benchmark values for the absolute size of the Yukawa couplings, because larger couplings would require $m_{\Delta}\gtrsim 3$~TeV to fulfill the experimental limits on CLFV processes. 
To quantify the corrections to the would-be zero entries, 
we first assume a given texture to be defined at the scale  $\Lambda_\text{UV}$ in the basis $Y_{\ell}(\Lambda_\text{UV})=\text{diag}\{y_e, y_{\mu}, y_{\tau}\}$.
The triplet mass $m_{\Delta}$ is taken equal to 3 TeV and the value of the cutoff scale is randomly varied in the range $3 \le \Lambda_\text{UV}\le 3\times10^{14}$~TeV. 
Then, we adopt the approximation that the flavor structure of $Y_{\Delta}$ varies slowly so that $Y_{\Delta ij}(\Lambda)=k Y_{\Delta ij}(m_{\Delta})$ if $Y_{\Delta ij}(\Lambda_\text{UV})\neq 0$, where $k$ is a common overall factor to be fixed. We also take $\left(Y_{\Delta}^{\dagger}Y_{\Delta}\right)_{\text{eff}}\approx Y_{\Delta}^{\dagger}(m_{\Delta})Y_{\Delta}(m_{\Delta})$ because the deviation from this approximation is always suppressed by an additional loop factor.
Next, we use the solution to the RG equations shown in Eq.~\eqref{sol1} to evolve $Y_\Delta, Y_{\ell}$ from $\Lambda_{\text{UV}}$ to $m_{\Delta}$.
Afterwards, we numerically diagonalize $Y_{\ell}(m_{\Delta})$ by a bi-unitary transformation $U_R^{\dagger}Y_{\ell}(m_{\Delta})U_L$, extract the unitary matrix $U_L$, and rotate $Y_{\Delta}(m_{\Delta})$ to the basis of the charged lepton mass eigenstates, $Y_{\Delta}(m_{\Delta})\to U_L^{T}Y_{\Delta}(m_{\Delta})U_L$. 
Finally, we rescale $Y_{\Delta}(m_{\Delta})$ by multiplying a new overall factor so that $Y_{\Delta\mu\tau}(m_{\Delta})$ matches the benchmark value $0.1$ or $0.3$, and extract the generated entry $Y_{\Delta\mu e}(m_{\Delta})$ for textures $\mathbf{A_1}$,  $\mathbf{B_2}$, $\mathbf{B_3}$, and $Y_{\Delta e e}(m_{\Delta})$ for textures \revised{$\mathbf{A_1}, \mathbf{A_2}$}. 

The top-left and bottom-left panels of Figure~\ref{RGE2} show that the ratios of BRs remain nearly constant, indicating that the $\mathbf{A_1}$, $\mathbf{A_2}$ textures are stable under the RG running as expected. Consequently, these textures can be consistently defined at all scales without invoking a UV completion to stabilize the zero entries. Their prediction for the ratio $\text{BR}(\mu\to \bar e ee)/\text{BR}(\mu\to e\gamma)$ thus serves as a robust signature to distinguish them from one another or from other textures. Falsifying these relations would directly exclude texture $\mathbf{A_1}$ or $\mathbf{A_2}$ within type~II seesaw, even without additional future information from neutrino oscillation experiments. 
In contrast, the top-central and top-right panels show that, for textures $\mathbf{B_2}$ and $\mathbf{B_3}$, the ratio $\text{BR}(\mu\to \bar e ee)/\text{BR}(\mu\to e\gamma)$ can be enhanced by more than two orders of magnitude and is sensitive to $\Lambda_\text{UV}$ if there is a large hierarchy between $m_{\Delta}$ and $\Lambda_\text{UV}$ and the elements in $Y_{\Delta}$ are sizable. The corresponding bottom panels also show that $\mu\to \bar e ee$ becomes more constraining than $\tau\to\bar \mu ee$ in such a case.
On the other hand, sizable RG effects do not exclude the possibility that $\tau\to\overline{\mu}ee$ still be observed first. This is the case even for $Y_{\Delta\mu\tau}\approx 0.3$ and a separation between $\Lambda_\text{UV}$ and $m_{\Delta}$ by up to four orders of magnitude.
This informs us that the zero-texture suppression of $\mu\to \bar e ee$ remains generally effective when sizable RG effects are taken into account. 
Consequently, although the $\mathbf{B_2}$, $\mathbf{B_3}$ textures require some kind of UV completion for stability, the corresponding UV physics can be decoupled in practice without inducing over-enhanced $\mu\to \bar e ee$ rates. This feature is crucial, as it ensures that contributions from $\Lambda_\text{UV}$-suppressed dimension-six operators can remain negligible. 

The results in Figure~\ref{RGE2} show that, in our scenario, RG effects may deteriorate texture stability and thus robust predictions and correlations. On the other hand, if that occurs it also opens up the interesting possibility of obtaining information on $\Lambda_\text{UV}$ by measuring ratios of CLFV branching ratios.  

\subsection{Information from colliders}
\label{sec:collider}

Determining $\Lambda_\text{UV}$ --- the scale at which an enhanced flavor \revised{symmetry} may emerge to generate or protect the texture zeros, typically far above the energy scales accessible to high-energy or high-intensity frontier experiments --- based on the above-discussed RG effects would require knowledge of the absolute strength of the triplet Yukawa couplings, that is, of  $Y_{\Delta\mu\tau}$ in our parameterization. 
Extracting this parameter is, in principle, possible at the LHC or future colliders \revised{by measuring the triplet mass and combining this piece of information with the coefficients of the dimension-six operators obtained from low-energy processes.}
For instance, if $\Delta$ lies not far above the TeV scale, it can be pair-produced at the LHC through the electroweak Drell-Yan process, $pp\to \Delta^{++}\Delta^{--}$ --- see e.g.~Ref.~\cite{Cai:2017mow, Ashanujjaman:2021txz}. These doubly-charged states are expected to decay dominantly to charged leptons since the $W^\pm W^\pm$ decay mode is suppressed by the small $L$-breaking parameter $\mu_\Delta$, leading to spectacular 4-lepton signatures, \mbox{$pp\to \Delta^{++}(\to \ell^+\ell^+) \Delta^{--} (\to \ell^-\ell^-)$}. 
The current lower bound $m_{\Delta}\gtrsim 1$~TeV~\cite{ATLAS:2022pbd} was obtained performing such kind of search and assuming equal BRs for the $\Delta^{\pm\pm}$ decays.
In addition, directly producing $\Delta^{\pm\pm}$ pairs even heavier than the TeV scale would be possible at a future high-energy muon collider~\cite{InternationalMuonCollider:2025sys} --- see~\cite{Li:2023ksw, Maharathy:2023dtp} for dedicated studies within the minimal type~II seesaw model --- or at the proposed $\sqrt{s} \approx 100$~TeV $pp$ colliders FCC-hh~\cite{FCC:2018vvp} and SppC~\cite{CEPCStudyGroup:2018rmc}.
In particular, $\Delta^{++}$ can also be singly produced via $\mu^+\mu^-\to \mu^-\ell^-\Delta^{++}$~\cite{Jia:2024wqi} at a $\mu^+\mu^-$ collider or through $\mu^+\mu^+\to \Delta^{++}(\gamma)$~\cite{Dev:2023nha, Das:2024kyk} at a $\mu^+\mu^+$ collider such as the $\mu$TRISTAN proposal~\cite{Hamada:2022mua}. Compared to pair production, these single production modes require a lower energy threshold so that producing $2-3$~TeV $\Delta^{++}$ could be feasible for near-term technology.

If $m_{\Delta}$ is measured as the location of a resonance peak, this would allow directly determining the Yukawa coupling $Y_{\Delta\mu\tau}$, once this observation is combined with information, e.g., on \mbox{$\text{BR}(\mu\to e\gamma)$}.\footnote{Even if $\Delta^{++}$ is too heavy to be produced on shell, high-$p_T$ observables, such as $\sigma(\mu^+\mu^+\to \ell^+_i\ell^+_j)$, can still be used to extract $m_{\Delta}$. The cross sections predicted in the full type~II seesaw model differ from those obtained from the dimension-six EFT (whose WCs can be inferred from the CLFV observables) by a factor of $\mathcal{O}(s/m_{\Delta}^2)$~\cite{Hamada:2022uyn, Fridell:2023gjx, Lichtenstein:2023iut}. At $\mu$TRISTAN with $\sqrt{s}=2$~TeV, such deviation is sizable even for $m_{\Delta}\simeq 4-5$ TeV.}
In principle, one could directly extract the Yukawa couplings by measuring the width of the resonance. However, the latter may be too narrow in realistic scenarios. Figure~\ref{scat} shows that $|Y_{\Delta\mu\tau}|\lesssim 0.2$ for $m_{\Delta}=3$~TeV, hence $|Y_{\Delta\mu\tau}|\lesssim 0.07$ for $m_{\Delta}=1$~TeV. As a consequence, the resonance width $\Gamma_\Delta \sim m_{\Delta} {Y_{\Delta \mu\tau}^2}/{(8\pi)}$ is at most $\approx 5$~GeV for $m_{\Delta}=3$~TeV ($\approx 0.2$~GeV for $m_{\Delta}=1$~TeV).
These numbers have to be compared with leptonic momentum resolutions. The relative resolution of LHC detectors for electrons with a $p_T$ between a few \revised{hundred}~GeV and 1~TeV is $\approx 1~\%$~\cite{CMS:2020uim}, while for muons it is as large as $\approx 5-10~\%$~\cite{CMS:2018rym} since, in the latter case, the deterioration of the tracker resolution cannot be compensated by electromagnetic calorimeter~(ECAL) information. Hence, combining collider information with CLFV observations seems to be a more feasible way to determine the absolute size of the Yukawa couplings and thus extract information on $\Lambda_\text{UV}$.

\begin{table}[t!]
\renewcommand\arraystretch{2.5}
\centering
\resizebox{1.\textwidth}{!}{
$\begin{array}{c|ccccccc}
\hline
\text{decay mode} & \bf{A_1}& \bf{A_2}& \bf{B_1} & \bf{B_2}& \bf{B_3}& \bf{B_4} & \bf{C} \\
 \hline
\Delta^{++}\to e^+\mu^+ & \text{---} & 8.0_{-1.0}^{+1.7}~\% & 0.25_{-0.23}^{+0.50}~\% & \text{---} & \text{---} & 0.16_{-0.15}^{+0.26}~\% & 12_{-12}^{+10}~\% \\
\Delta^{++}\to e^+\tau^+ & 8.3_{-1.0}^{+1.8}~\%  & \text{---} & \text{---} & 0.25_{-0.24}^{+0.50}~\%  & 0.16_{-0.14}^{+0.26}~\%  & \text{---} & 9_{-9}^{+12}~\% 
   \\
\Delta^{++}\to \mu^+\tau^+  & 39_{-5}^{+8}~\%  & 41_{-6}^{+9}~\%  & 69.8_{-1.5}^{+0.3}~\%  & 69.8_{-1.7}^{+0.3}~\%  & 69.8_{-1.7}^{+0.4}~\%  &
   69.8_{-1.9}^{+0.4}~\%  & 49_{-24}^{+23}~\%  \\
\Delta^{++}\to e^+e^+ & \text{---} & \text{---} & 28_{-4}^{+4}~\%  & 28_{-4}^{+4}~\% & 28_{-4}^{+4}~\% & 28_{-4}^{+4}~\% & 30_{-20}^{+12}~\% \\
\Delta^{++}\to \mu^+\mu^+  & 27_{-8}^{+10}~\% & 27_{-7}^{+5}~\% & \text{---} & 2.5_{-2.3}^{+4.0}~\% & \text{---} & 2.1_{-1.9}^{+3.5}~\% & \text{---} \\
\Delta^{++}\to \tau^+\tau^+  & 25_{-7}^{+5}~\% & 24_{-8}^{+11}~\% & 2.5_{-2.2}^{+4.0}~\% & \text{---} & 2.0_{-1.8}^{+3.5}~\% & \text{---} & \text{---}  \\
\hline
\end{array}$}
    \caption{Central values and $3\sigma$ uncertainties for the branching ratios of $\Delta^{++} \to \ell_i^+\ell_j^{+}$ decays.} 
    \label{branchingRatios}
\end{table}

On the other hand, if $\Delta^{\pm\pm}$ is observed at colliders, its decay BRs also contain information on the relative pattern of $Y_\Delta$ and can complement CLFV data. We calculate the 6 BRs of $\Delta^{++} \to \ell^+_i \ell^+_j$ for the 7 two-zero textures using the generated $10^4$ possible configurations of $Y_{\Delta}$ and show the results in Table~\ref{branchingRatios}. As for Table~\ref{3body}, the displayed $3\sigma$ uncertainties correspond to the range that 99.7 \% of the output numbers lie within. The dash indicates the vanishing decay modes for each pattern. We do not include RG running effects for $Y_{\Delta}$ as they would only give subdominant corrections to these observables without modifying the overall patterns.
As one can see from the table, for all textures, the dominant decay mode is typically $\Delta^{++}\to \mu^+\tau^+$. Besides that, each texture features a distinctive pattern of $\Delta^{++}$ decays. Hence, the (lucky) observation of $\Delta^{++}$ at colliders would allow a direct test of the existence of this kind of flavor structures. 
This result is to be compared with the predictions from generic textures, which are subject to large uncertainties from the unknown absolute mass scale and Majorana phases~\cite{Garayoa:2007fw, FileviezPerez:2008jbu, Cai:2017mow}.
In addition, high-$p_T$ processes such as $\mu^+\mu^+\to \ell^+_i\ell^+_j$ can also provide information on the flavor structure of the matrix $Y_\Delta$~\cite{Fridell:2023gjx}, in a way similar to the BRs of Table~\ref{branchingRatios}. However, this probe is only possible for textures $\mathbf{A_1}, \mathbf{A_2}, \mathbf{B_2},$ and $ \mathbf{B_4}$, since $Y_{\Delta\mu\mu}$ vanishes for the others.


\section{Conclusions}
\label{conclu}

Working in the context of the minimal type~II seesaw, the results of this paper demonstrate that the two-zero textures originally imposed on the Majorana neutrino mass matrix $M^{\nu}$ can be consistently extended to the flavor structure of NP at a scale as low as $5-10$ TeV, leading to distinctive correlations among several CLFV transition rates. 
The results of our phenomenological study presented in Section~\ref{phenosection} provide a further example of the model discriminating power of CLFV searches and of their capability to shed light on the NP flavor structure, in particular in the fortunate eventuality that more than one CLFV process is observed in the ongoing experimental campaign.

Compared with the flavor patterns protected by symmetries such as $U(2)$, $A_4$, and $Z_3$, the two-zero textures that we studied are equally predictive with respect to the following key aspects:
\begin{enumerate}[label=\roman*]
    \item Certain two-zero textures can suppress the dangerous $\mu\to e$ transitions and allow a relatively low effective cutoff scale. NP at a few TeV can thus remain compatible with sizable Yukawa couplings.
    \item Several CLFV processes can simultaneously lie within the reach of future experiments. The ratios among their BRs are predicted, providing a possible method to distinguish some of the two-zero textures from each other and the other flavor structures.
    \item Although not all two-zero textures are stable under RG evolution, $\mu\to e$ transitions could remain suppressed even after the RG effects are included. By quantifying these corrections, the CLFV observables --- together with other TeV scale probes --- can potentially provide information on the ultra-high scale related to the origin of two-zero textures.
\end{enumerate}

In addition, we think that the physical implications of our work extend beyond seesaw mechanisms and zero textures. 
For decades, precision measurements on processes such as $\mu\to e\gamma$ and neutral $K$ meson oscillations have pushed the possible effective scale of transitions between the first and second generation charged fermions to very high energies. Consequently, it is commonly believed that TeV scale NP is only possible if accompanied with a protection mechanism from flavor symmetries in order to evade these stringent constraints. So far, the landscape of new renormalizable models or effective theories with MFV, $U(2)$, or discrete symmetries such as $A_4$ and $Z_3$ has been studied in depth. 
However, our examples suggest that simple and predictive benchmarks can emerge even when the underlying flavor symmetries (if any exist) remain implicit. We therefore argue that what we might call the \textit{swampland of flavor symmetries} could exhibit rich and unexplored structures for phenomenology in general. 

\section*{Acknowledgments}
We would like to thank Marco Ardu, Saiyad Ashanujjaman, Di Zhang, Shun Zhou, and Xunwu Zuo for valuable discussions.
This research was partially supported by the National Natural Science Foundation of China (NSFC)
under grant No.~12035008 and the Deutsche Forschungsgemeinschaft (DFG, German Research Foundation) under grant 396021762 - TRR 257. X.G.~also acknowledges support by the Doctoral School ``Karlsruhe School of Elementary and Astroparticle Physics: Science and Technology''.


\section*{Appendix}
\appendix

\section{Penguin diagram loop function}
\label{loopfunction}
The complete form of the loop function $f\left(\frac{-q^2}{m_{\Delta}^2},\frac{m_j^2}{m_{\Delta}^2}\right)$ appearing in Eq.~\eqref{eftWilson} reads~\cite{Raidal:1997hq}:
\begin{equation}
\resizebox{0.93\hsize}{!}{%
    $f\left(\frac{-q^2}{m_{\Delta}^2},\frac{m_j^2}{m_{\Delta}^2}\right)~=~\log\left(\frac{m_j^2}{m_{\Delta}^2}\right)-\frac{4 m_j^2}{q^2}+\left(\left(1+\frac{2 m_j^2}{q^2}\right)\sqrt{1-\frac{4 m_j^2}{q^2}} \log\left( \frac{\sqrt{1-\frac{4 m_j^2}{q^2}}+1}{\sqrt{1-\frac{4 m_j^2}{q^2}}-1}\right)\right)\,.$%
    }
\end{equation}
While Eq.~\eqref{loopf} is a good approximation when $-q^2\lesssim m_j^2$, the threshold correction $5/3$ is absent if $-q^2\gg m_j^2$:
\begin{equation}
    f\left(\frac{-q^2}{m_{\Delta}^2},\frac{m_j^2}{m_{\Delta}^2}\right)~=~ \log\left(\frac{-q^2}{m_{\Delta}^2}\right), \quad \text{when }  -q^2\gg m_k^2\,. 
\end{equation}
This corresponds to the electron loop contributing to the $\mu\rightarrow \bar e  ee $ decay.

\section{Explicit two-zero textures}
\label{expressions}
For the textures $\mathbf{A_1}$, $\mathbf{B_1}$, $\mathbf{B_3}$, $\mathbf{C}$ in Eq.~\eqref{twozero}, the flavor structure of $M^{\nu}$ is related to the neutrino mixing angles $\theta_{13}$, $\theta_{12}$, $\theta_{12}$ and the Dirac CPV phase $\delta$ through the following expressions. Here, we simplify the results for $\mathbf{B_1}$, $\mathbf{B_3}$ presented in Ref.~\cite{Kitabayashi:2015jdj} and generalize them to textures $\mathbf{A_1}$ and $\mathbf{C}$.

\vspace{10pt}
\noindent
$\mathbf{A_1}$:
\begin{equation}
\label{A1f}
\begin{aligned}
    \frac{y}{a}~=~&-e^{i \delta } \cos ^3\left(\theta _{23}\right) \tan \left(\theta _{13}\right)+2 e^{-i \delta } \sin ^2\left(\theta
   _{23}\right) \cos \left(\theta _{23}\right) \tan \left(\theta _{13}\right)\\&+2 e^{-i \delta } \sin ^2\left(\theta
   _{23}\right) \cos \left(\theta _{23}\right) \cot \left(2 \theta _{13}\right)-\sin \left(2 \theta _{23}\right)
   \cos \left(\theta _{23}\right) \cot \left(2 \theta _{12}\right) \sec \left(\theta _{13}\right)\,,\\
   \frac{c}{a}~=~&\frac{1}{2} e^{i \delta } \sin \left(2 \theta _{23}\right) \cos \left(\theta _{23}\right) \tan \left(\theta
   _{13}\right)+e^{-i \delta } \sin \left(\theta _{23}\right) \cos \left(2 \theta _{23}\right) \tan \left(\theta
   _{13}\right)\\&+e^{-i \delta } \sin \left(2 \theta _{23}\right) \cos \left(\theta _{23}\right) \cot \left(2 \theta
   _{13}\right)+\sin \left(\theta _{23}\right) \sin \left(2 \theta _{23}\right) \cot \left(2 \theta _{12}\right)
   \sec \left(\theta _{13}\right)\,,\\
   \frac{z}{a}~=~& \frac{3}{2} e^{-i \delta } \cos \left(\theta _{23}\right) \cot \left(2 \theta _{13}\right)+\frac{1}{2} e^{-i \delta
   } \cos \left(3 \theta _{23}\right) \cot \left(2 \theta _{13}\right)\\&-2 e^{-i \delta } \sin ^2\left(\theta
   _{23}\right) \cos \left(\theta _{23}\right) \tan \left(\theta _{13}\right)-e^{i \delta } \sin ^2\left(\theta
   _{23}\right) \cos \left(\theta _{23}\right) \tan \left(\theta _{13}\right)\\&-2 \sin ^3\left(\theta _{23}\right)
   \cot \left(2 \theta _{12}\right) \sec \left(\theta _{13}\right)\,,
\end{aligned}
\end{equation}

\vspace{10pt}
\noindent
$\mathbf{B_1}$:
\begin{equation}
\label{B1f}
\begin{aligned}
\frac{x}{b}~=~& \left(-4 e^{i \delta } \sin ^3\left(\theta _{23}\right) \cot \left(2 \theta _{13}\right)+4 e^{i \delta } \sin \left(\theta
   _{23}\right) \cos ^2\left(\theta _{23}\right) \tan \left(\theta _{13}\right)\right.\\&\left.+2 e^{3 i \delta } \sin \left(\theta
   _{23}\right) \cos ^2\left(\theta _{23}\right) \tan \left(\theta _{13}\right)-4 e^{2 i \delta } \cos
   ^3\left(\theta _{23}\right) \cot \left(2 \theta _{12}\right) \sec \left(\theta _{13}\right)\right)
   \\& \times \left(\sin ^2\left(\theta _{23}\right)+e^{2 i \delta } \cos ^2\left(\theta _{23}\right)\right)^{-1}\,,\\
   \frac{c}{b}~=~& \left( \frac{1}{2} e^{-i \delta } \cos \left(\theta _{23}\right) \tan \left(\theta _{13}\right)-\frac{1}{2} e^{i \delta }
   \cos \left(\theta _{23}\right) \tan \left(\theta _{13}\right)\right.\\&\left.-\frac{1}{2} e^{-i \delta } \cos \left(\theta
   _{23}\right) \cos \left(2 \theta _{23}\right) \tan \left(\theta _{13}\right)-\frac{1}{2} e^{i \delta } \cos
   \left(\theta _{23}\right) \cos \left(2 \theta _{23}\right) \tan \left(\theta _{13}\right)\right.\\&\left.+e^{i \delta } \sin
   ^2\left(\theta _{23}\right) \cos \left(\theta _{23}\right) \cot \left(\theta _{13}\right)-\sin \left(2 \theta
   _{23}\right) \cos \left(\theta _{23}\right) \cot \left(2 \theta _{12}\right) \sec \left(\theta _{13}\right)\right)
   \\& \times \left(\sin ^2\left(\theta _{23}\right)+e^{2 i \delta } \cos ^2\left(\theta _{23}\right)\right)^{-1}\,,\\
   \frac{z}{b}~=~& \left( \frac{1}{2} e^{-i \delta } \tan \left(\theta _{13}\right) \csc \left(\theta _{23}\right)+\frac{1}{2} e^{i \delta }
   \tan \left(\theta _{13}\right) \csc \left(\theta _{23}\right)\right.\\&\left.-\frac{1}{2} e^{-i \delta } \cos ^2\left(2 \theta
   _{23}\right) \tan \left(\theta _{13}\right) \csc \left(\theta _{23}\right)-\frac{1}{2} e^{i \delta } \cos
   ^2\left(2 \theta _{23}\right) \tan \left(\theta _{13}\right) \csc \left(\theta_{23}\right)\right.\\&\left.+e^{i \delta } \sin
   \left(\theta _{23}\right) \cos \left(2 \theta _{23}\right) \cot \left(\theta _{13}\right)-\frac{1}{2} \sin
   \left(4 \theta _{23}\right) \cot \left(2 \theta _{12}\right) \csc \left(\theta _{23}\right) \sec \left(\theta
   _{13}\right)\right)
   \\& \times \left(\sin ^2\left(\theta _{23}\right)+e^{2 i \delta } \cos ^2\left(\theta _{23}\right)\right)^{-1}\,,
\end{aligned}
\end{equation}

\vspace{10pt}
\noindent
$\mathbf{B_3}$:
\begin{equation}
\label{B3f}
\begin{aligned}
\frac{x}{a}~=~& \left(2 e^{3 i \delta } \cos ^3\left(\theta _{23}\right) \tan \left(\theta _{13}\right)-e^{i \delta } \cos \left(\theta
   _{23}\right) \cot \left(2 \theta _{13}\right)+e^{i \delta } \cos \left(3 \theta _{23}\right) \cot \left(2 \theta
   _{13}\right)\right.\\&\left.-4 e^{i \delta } \sin ^2\left(\theta _{23}\right) \cos \left(\theta _{23}\right) \tan \left(\theta
   _{13}\right)+2 e^{2 i \delta } \sin \left(2 \theta _{23}\right) \cos \left(\theta _{23}\right) \cot \left(2
   \theta _{12}\right) \sec \left(\theta _{13}\right)\right)
   \\& \times \left(\sin ^2\left(\theta _{23}\right)+e^{2 i \delta } \cos ^2\left(\theta _{23}\right)\right)^{-1}\,,\\
   \frac{c}{a}~=~& \left( -e^{-i \delta } \sin \left(\theta _{23}\right) \tan \left(\theta _{13}\right)+e^{i \delta } \sin \left(\theta
   _{23}\right) \tan \left(\theta _{13}\right)\right.\\&\left.+2 e^{i \delta } \sin \left(\theta _{23}\right) \cos ^2\left(\theta
   _{23}\right) \cot \left(\theta _{13}\right)+e^{-i \delta } \sin \left(\theta _{23}\right) \cos \left(2 \theta
   _{23}\right) \tan \left(\theta _{13}\right)\right.\\&\left.+e^{i \delta } \sin \left(\theta _{23}\right) \cos \left(2 \theta
   _{23}\right) \tan \left(\theta _{13}\right)+2 \sin \left(\theta _{23}\right) \sin \left(2 \theta _{23}\right)
   \cot \left(2 \theta _{12}\right) \sec \left(\theta _{13}\right) \right)
   \\& \times \left(\sin ^2\left(\theta _{23}\right)+e^{2 i \delta } \cos ^2\left(\theta _{23}\right)\right)^{-1}\,,\\
   \frac{z}{a}~=~& \left( e^{i \delta } \cos \left(\theta _{23}\right) \cot \left(\theta _{13}\right)+e^{i \delta } \cos \left(3 \theta
   _{23}\right) \cot \left(\theta _{13}\right)-e^{-i \delta } \tan \left(\theta _{13}\right) \sec \left(\theta
   _{23}\right)\right.\\&\left.-e^{i \delta } \tan \left(\theta _{13}\right) \sec \left(\theta _{23}\right)+e^{-i \delta } \cos
   ^2\left(2 \theta _{23}\right) \tan \left(\theta _{13}\right) \sec \left(\theta _{23}\right)\right.\\&\left.+e^{i \delta } \cos
   ^2\left(2 \theta _{23}\right) \tan \left(\theta _{13}\right) \sec \left(\theta _{23}\right)+\sin \left(4 \theta
   _{23}\right) \cot \left(2 \theta _{12}\right) \sec \left(\theta _{13}\right) \sec \left(\theta _{23}\right)\right)
   \\& \times \left(\sin ^2\left(\theta _{23}\right)+e^{2 i \delta } \cos ^2\left(\theta _{23}\right)\right)^{-1}\,,
\end{aligned}
\end{equation}

\vspace{10pt}
\noindent
$\mathbf{C}$:
\begin{equation}
\label{cf}
    \begin{aligned}
    \frac{x}{c}~=~&e^{2 i \delta } \tan ^2\left(\theta _{13}\right) \csc \left(2 \theta _{23}\right)-e^{2 i \delta } \cos \left(2
   \theta _{23}\right) \tan ^2\left(\theta _{13}\right) \cot \left(2 \theta _{23}\right)\\&+2 e^{i \delta } \cos
   \left(2 \theta _{13}\right) \cos \left(2 \theta _{23}\right) \cot \left(2 \theta _{12}\right) \csc \left(\theta
   _{13}\right) \sec ^2\left(\theta _{13}\right)-\csc \left(2 \theta _{23}\right)\\&+\cos \left(2 \theta _{23}\right)
   \cot \left(2 \theta _{23}\right)+\tan ^2\left(\theta _{13}\right) \csc \left(2 \theta _{23}\right)-\cos \left(2
   \theta _{23}\right) \tan ^2\left(\theta _{13}\right) \cot \left(2 \theta _{23}\right)\,,\\
\frac{b}{c}~=~& -\frac{1}{2} e^{i \delta } \cos \left(\theta _{23}\right) \cot \left(\theta _{13}\right)-\frac{1}{2} e^{i \delta }
   \cos \left(3 \theta _{23}\right) \cot \left(\theta _{13}\right)+\frac{1}{2} e^{-i \delta } \tan \left(\theta
   _{13}\right) \sec \left(\theta _{23}\right)\\&+\frac{1}{2} e^{i \delta } \tan \left(\theta _{13}\right) \sec
   \left(\theta _{23}\right)-e^{-i \delta } \sin \left(\theta _{23}\right) \cos \left(2 \theta _{23}\right) \tan
   \left(\theta _{13}\right) \cot \left(2 \theta _{23}\right)\\&-e^{i \delta } \sin \left(\theta _{23}\right) \cos
   \left(2 \theta _{23}\right) \tan \left(\theta _{13}\right) \cot \left(2 \theta _{23}\right)\\&-2 \sin \left(\theta
   _{23}\right) \cos \left(2 \theta _{23}\right) \cot \left(2 \theta _{12}\right) \sec \left(\theta _{13}\right)\,,\\
   \frac{a}{c}~=~& -\frac{1}{2} e^{i \delta } \sin \left(\theta _{23}\right) \cot \left(\theta _{13}\right)+\frac{1}{2} e^{i \delta }
   \sin \left(3 \theta _{23}\right) \cot \left(\theta _{13}\right)\\&+2 e^{-i \delta } \sin \left(\theta _{23}\right)
   \cos ^2\left(\theta _{23}\right) \tan \left(\theta _{13}\right)+2 e^{i \delta } \sin \left(\theta _{23}\right)
   \cos ^2\left(\theta _{23}\right) \tan \left(\theta _{13}\right)\\&-2 \cos \left(2 \theta _{23}\right) \cos
   \left(\theta _{23}\right) \cot \left(2 \theta _{12}\right) \sec \left(\theta _{13}\right)\,.
    \end{aligned}
\end{equation}

To calculate the possible values for the entries of $Y_{\Delta}$ and analyze their \revised{distributions}, we first generate $10^4$ points of the parameter sets $\{\theta_{13},\,\theta_{12},\, \theta_{12},\,\Delta m^2_{12},\,\Delta m^2_{13}\}$ that satisfy the distributions resulting from the fit to neutrino oscillation data reported in~\cite{Esteban:2024eli}. 
$M^{\nu}$ for texture $\mathbf{A_1}$, $\mathbf{B_1}$, $\mathbf{B_3}$ and $\mathbf{C}$ are then calculated with Eq.~\eqref{A1f}-\eqref{cf}, using the updated parameter set $\{\theta_{13},\, \theta_{12},\, \theta_{12},\, \delta\}$. 
\revised{
The CP violating phase $\delta$ is not a free parameter but is determined by the three mixing angles and the ratio $\Delta m^2_{12}/|\Delta m^2_{13}|$, whose explicit expressions are reported in Ref.~\cite{Fritzsch:2011qv}. 
This relation fully fixes the neutrino mass matrices with texture zeros, rendering the neutrino mass ordering a prediction~\cite{Meloni:2014yea,Zhou:2015qua, Treesukrat:2025dhd}. 
The $\textbf{A}$ textures always lead to the NO mass spectrum.
For $\textbf{B}$ textures, the mass ordering depends on the octant of $\theta_{23}$. 
$\theta_{23}<45^{\circ}$ yields the NO spectrum for $\mathbf{B_1}$, $\mathbf{B_3}$ and IO spectrum $\mathbf{B_2}, \mathbf{B_4}$. 
Conversely, $\theta_{23}>45^{\circ}$ generates the dual solution, IO spectrum for $\mathbf{B_1}$, $\mathbf{B_3}$ and NO spectrum $\mathbf{B_2}, \mathbf{B_4}$.
We have checked that these two scenarios lead to nearly the same phenomenological consequences. 
The texture $\textbf{C}$ produces the NO spectrum only for $\theta_{23}=45^{\circ}$; otherwise, it predicts the IO spectrum.}

As a result of the $\mu-\tau$ reflection symmetry~\cite{Harrison:2002et, Xing:2015fdg, Xing:2022uax}, textures $\mathbf{A_2}$, $\mathbf{B_2}$, $\mathbf{B_4}$ are dual to textures $\mathbf{A_1}$, $\mathbf{B_1}$, $\mathbf{B_3}$ and correlated as follows: 
\begin{equation}
\label{dual}
a \leftrightarrow b^*\,,~~ y\leftrightarrow z^*\,,~~ \theta_{23}\leftrightarrow \frac{\pi}{2}-\theta_{23}\,,~~\delta\leftrightarrow \delta-\pi\,,
\end{equation}
up to unphysical charged lepton phases. The relations for the flavor structures $\mathbf{A_2}$, $\mathbf{B_2}$, $\mathbf{B_4}$ can then be obtained by substituting Eq.~\eqref{dual} into Eq.~\eqref{A1f}-\eqref{B3f}. 

Since $Y_{\Delta}$ is proportional to $M^{\nu}$, the above formulae enable us to express the entries of the matrix $Y_{\Delta}$ in terms of physically observable parameters.
As an illustration, we fix $Y_{\Delta\mu\tau}=1$ and provide one numerical benchmark of the flavor structure of $|Y_{\Delta ij}|$ for each texture that is compatible with neutrino oscillation data within the current uncertainties: 
\begin{equation}
\label{twozerobenchmark}
\begin{aligned}
\mathbf{A_1} &: \left(
\begin{array}{ccc}
 0 & 0 & 0.43 \\
 0 & 1.07 & 1\\
 0.43 & 1& 1.12 \\
\end{array}
\right),~  \mathbf{A_2}: \left(
\begin{array}{ccc}
 0 & 0.45 & 0 \\
 0.45 & 1.15 & 1\\
 0 & 1& 1.09 \\
\end{array}
\right),
~\mathbf{B_1}: \left(
\begin{array}{ccc}
 0.83 & 0.10 & 0 \\
 0.10 & 0 & 1\\
 0 & 1& 0.42 \\
\end{array}
\right),\\ \mathbf{B_2} &: \left(
\begin{array}{ccc}
 0.87 & 0 & 0.07 \\
 0 & 0.31 & 1\\
 0.07 & 1& 0 \\
\end{array}
\right),
~ \mathbf{B_3} : \left(
\begin{array}{ccc}
 0.84 & ~~0~~ & 0.07 \\
 0 & 0 & 1\\
 0.07 & 1& 0.38 \\
\end{array}
\right),
~\mathbf{B_4} :\left(
\begin{array}{ccc}
 0.87 & 0.06 & 0 \\
 0.06 & 0.29 & 1\\
 0 & 1& ~~0~~ \\
\end{array}
\right),\\
\mathbf{C^{~}_{~}}&: \left(
\begin{array}{ccc}
 0.74 & 0.06 & 0.40 \\
 0.06 & 0 & 1\\
 0.40 & 1& 0 \\
\end{array}
\right).
\end{aligned}
\end{equation}
Notice that all the above two-zero textures break all $U(1)_{\ell}$, $U(2)_{\ell}$, $Z_3$, or $A_4$ flavor symmetries.

\section{More details on RG equations}
\label{overall}

The overall factors in Eq.~\eqref{sol0} and Eq.~\eqref{sol1} read:
\begin{equation}
\label{Idef}
\begin{aligned}
    I_{g\lambda t}~&=~\left(\frac{\mu}{m_{\Delta}}\right)^{\frac{1}{16\pi^2}\left(-3 g_2^2+2 \lambda_H+6y_t^2 \right)_{\text{eff}}}, \qquad
    I_{\ell}~=~ \left(\frac{\mu}{m_{\Delta}}\right)^{-\frac{1}{32\pi^2}\left(y_{\ell}^2\right)_{\text{eff}}},\\
    I_{tg}~&=~ \left(\frac{\Lambda_\text{UV}}{m_{\Delta}}\right)^{\frac{1}{16\pi^2}\left(3y_t^2- \frac{9}{4}g_2^2- \frac{9}{4}g_1^2 \right)_{\text{eff}}}, \qquad
    I_{\Delta g}~=~ \left(\frac{\Lambda_\text{UV}}{m_{\Delta}}\right)^{\frac{1}
    {16\pi^2}\left( \text{Tr} \left[Y_{\Delta}^{\dagger}Y_{\Delta}\right]-\frac{9}{10}g_1^2-\frac{9}{2}g_2^2 \right)_{\text{eff}}},\\
\end{aligned}
\end{equation}
where the explicit definition of the effective coupling matrix $\left(Y_{\Delta}^{\dagger}Y_{\Delta}\right)_{\text{eff}}$ is given by
\begin{equation}
\label{Ydef}
\left(Y_{\Delta}^{\dagger}Y_{\Delta}\right)_{\text{eff}}~=~\frac{\int^{\log{m_{\Delta}}}_{\log{\Lambda_\text{UV}}} \left(Y_{\Delta}^{\dagger}(\mu')Y_{\Delta}(\mu')\right) d\log{\mu'}}{\log{\left(m_{\Delta}/\Lambda_\text{UV}\right)}}, 
\end{equation}
and other effective couplings are defined analogously. However, since they vary slowly within a perturbative theory and always appear together with the loop factor $1/(16\pi^2)$, we replace the effective couplings with their values at $m_{\Delta}$ as an approximation. 

The matrix functions in Eq.~\eqref{sol1} are defined as: 
\begin{equation}
\label{matrixpower}
     \left(\frac{m_{\Delta}}{\Lambda_\text{UV}}\right)^{\frac{3}{16\pi^2}\left(Y_{\Delta}^{\dagger}Y_{\Delta}\right)_{\text{eff}}}~=~\exp{\left[\frac{3}{16\pi^2}\log{(m_{\Delta}/\Lambda_{\text{UV}})}\left(Y_{\Delta}^{\dagger}Y_{\Delta}\right)_{\text{eff}}\right] }.
\end{equation}
Here, $\exp{M}\equiv\mathbb{1}+M+\frac{1}{2!}M\cdot M+...$ can be evaluated as a matrix power expansion when $M$ is a square matrix. 

\bibliographystyle{JHEP} 
\bibliography{TypeII_2-0.bib}

\end{document}